\documentclass[twocolumn,showpacs,nofootinbib,preprintnumbers,amsmath,amssymb,showkeys,superscriptaddress]{revtex4-1}
\usepackage{graphicx}
\usepackage{amsmath}
\usepackage{amssymb}
\usepackage{dcolumn}% Align table columns on decimal point
\usepackage{bm}% bold math
\usepackage{color}
\usepackage{dsfont}
\usepackage{epsfig}

\newcommand \beq{\begin{eqnarray}}
\newcommand \eeq{\end{eqnarray}}

\begin{document}
\allowdisplaybreaks

\title{A Gribov-Zwanziger type model action invariant\\ under background gauge transformations}

\author{Daniel Kroff}
\email{daniel.kroff-fogaca@polytechnique.edu}
\affiliation{Centre de Physique Th\'eorique, Ecole Polytechnique, CNRS, Universit\'e Paris-Saclay, F-91128 Palaiseau, France.}

\author{Urko Reinosa}
\email{urko.reinosa@polytechnique.edu}
\affiliation{Centre de Physique Th\'eorique, Ecole Polytechnique, CNRS, Universit\'e Paris-Saclay, F-91128 Palaiseau, France.}

\date{\today}

\begin{abstract}
We propose a Gribov-Zwanziger type model action for the Landau-DeWitt gauge that preserves, for any gauge group, the invariance under background gauge transformations. At zero temperature, and to one-loop accuracy, the model can be related to the Gribov no-pole condition. We apply the model to the deconfinement transition in SU(2) and SU(3) Yang-Mills theories and compare the predictions obtained with a single or with various (color dependent) Gribov parameters that can be introduced in the action without jeopardizing its background gauge invariance. The Gribov parameters associated to color directions orthogonal to the background can become negative, while keeping the background effective potential real. In some cases, the proper analysis of the transition requires the potential to be resolved in those regions.
\end{abstract}

\maketitle

%%%%%
\section{Introduction}

Much progress has been achieved lately in the continuum description of the dynamics at play in the deconfinement transition of pure Yang-Mills theories. First, a good handle on the related center symmetry was possible thanks to the use of background field methods \cite{Abbott:1980hw,Abbott:1981ke} which allow for the definition of order parameters equivalent to the Polyakov loop, but simpler to compute in practice \cite{Braun:2007bx}. Second, relevant dynamics could be captured thanks to the use of sophisticated non-perturbative methods such as the functional renormalization group \cite{Braun:2007bx,Braun:2009gm,Braun:2010cy}, the infinite tower of Dyson-Schwinger equations \cite{Fischer:2009wc,Fischer:2009gk,Fischer:2011mz,Fischer:2013eca,Fischer:2014vxa} or variational approaches \cite{Reinhardt:2012qe,Reinhardt:2013iia,Quandt:2016ykm,Reinhardt:2017pyr}. 

On top of these achievements, more phenomenological approaches \cite{Tissier:2010ts,Tissier:2011ey,Reinosa:2014ooa,Canfora:2015yia} seem to indicate that, in the Landau gauge (and in its background extension, the so-called Landau-DeWitt gauge), a pivotal part of the dynamics may become accessible to perturbative methods, but only after a complete gauge-fixing procedure has been achieved, including the proper handling of the associated Gribov copy problem \cite{Gribov77}. In fact, according to these studies, once such a gauge-fixing is implemented, at least in some approximate form, the perturbative expansion becomes viable at low energies \cite{Tissier:2011ey,Reinosa:2017qtf}, while it breaks down in the more standard Faddeev-Popov gauge-fixing. This is an interesting perspective that could open the way to the perturbative evaluation of quantities that are usually considered as genuinely non-perturbative. Although speculative, the idea certainly deserves to be further investigated and tested. 

For instance, in a series of recent works, the Curci-Ferrari (CF) action \cite{Curci76} has been proposed as a model for a complete gauge-fixing in the Landau gauge \cite{Tissier:2010ts,Tissier:2011ey,Serreau:2012cg}. The underlying conjecture of these studies is that a CF gluon mass term may arise after the Gribov copies have been accounted for by means of an uneven averaging procedure \cite{Serreau:2012cg}. Although no rigorous mechanism for the generation of such a CF mass has been identified in the Landau gauge, a similar mass term could be generated in a non-linear version of the Landau gauge \cite{Tissier:2017fqf}. Moreover and interestingly, relatively simple one-loop calculations of zero-temperature correlation functions in the CF model \cite{Tissier:2010ts,Tissier:2011ey,Pelaez:2013cpa} agree pretty nicely with first principle lattice simulations of Yang-Mills correlation functions in the Landau gauge \cite{Cucchieri_08b,Cucchieri_08,Bornyakov2008,Cucchieri09,Bogolubsky09,Bornyakov09}. The model has also been extended to finite temperature, within the Landau-DeWitt gauge framework, where it gives a good description of center-symmetry breaking in pure Yang-Mills theories \cite{Reinosa:2014ooa}. In this case, two-loop corrections could also be computed \cite{Reinosa:2014zta,Reinosa:2015gxn}, showing some sign of apparent convergence and supporting the idea that perturbation theory may indeed be applicable once the Gribov problem has been properly handled. Finally, matter fields can also be included in the analysis, see Refs.~\cite{Pelaez:2014mxa,Pelaez:2015tba,Reinosa:2015oua,Maelger:2017amh}.

Another possible way to deal with the Gribov problem in the Landau gauge is the so-called Gribov-Zwanziger approach \cite{Gribov77,Zwanziger89,Vandersickel:2012tz}. The idea in that case is to restrict the domain of the functional integral to a region that contains less Gribov copies, in practice the so-called first Gribov region, defined by the positivity of the Faddeev-Popov operator $-\partial_\mu D_\mu$. With the price of introducing some auxiliary fields, a formulation of this restriction was constructed in terms of a local and renormalizable quantum field theory \cite{Zwanziger89}. It has since then known various refinements in order to match lattice results at zero-temperature \cite{Dudal08,Vandersickel:2011zc}.

At finite temperature, the situation is less clear. Although many interesting works apply the Gribov-Zwanziger approach to thermal scenarios \cite{Zwanziger:2004np,Zwanziger:2006sc,Fukushima:2013xsa,Su:2014rma,Florkowski:2015rua,Florkowski:2015dmm}, they all rely on the implicit assumption that the output of the Gribov-Zwanziger construction in such cases is given by the zero temperature Gribov-Zwanziger action taken over a compact (imaginary) time interval of length $\smash{\beta=1/T}$. Although natural, this assumption is far from obvious. In fact, as recently discussed in Refs.~\cite{Cooper:2015sza,Cooper:2015bia}, the presence of the compact time direction and the related periodic boundary conditions lift the degeneracy of the lowest, non-zero eigenvalues of the free Faddeev-Popov operator. This, in turn, leads to a modification of the Faddeev-Popov action which is not just the usual zero-temperature modification taken over a compact time interval. This approach certainly opens a new line of investigation towards a proper discussion of the Gribov-Zwanziger gauge-fixing at finite temperature in the Landau gauge. However, it also poses new questions. In particular, the so-obtained action is not invariant under O(4) Euclidean space-time rotations\footnote{These are the counterpart of Lorentz transformations in the imaginary time formalism.} in the zero-temperature limit, unless the Gribov parameter goes to zero. It is therefore not clear whether or how the model is renormalizable. Another issue is that, for the approach to correspond to a bona-fide gauge-fixing in the Landau gauge, the $O(4)$ breaking terms in the zero-temperature limit should not affect the physical observables. This question deserves further investigation and probably requires the identification of the appropriate BRST (Becchi-Rouet-Stora-Tyutin) symmetry.
  
In the case of the Landau-DeWitt gauge, the situation is similar to that of the Landau gauge prior to the results of Refs.~\cite{Cooper:2015sza,Cooper:2015bia}. There is to date no first principle derivation of the associated Gribov-Zwanziger action, only models that try to incorporate the effect of restricting the functional integral to the corresponding first Gribov region. In particular, in Ref.~\cite{Canfora:2015yia}, a Gribov-Zwanziger type action for the Landau-DeWitt gauge has been proposed -- independently of whether it corresponds to a faithful implementation of the Gribov restriction -- and applied to the study of center-symmetry breaking in SU(2) Yang-Mills theory (see also Ref.~\cite{Canfora:2016ngn}). This action has the convenient property that it reproduces the usual, renormalizable, $O(4)$ invariant, Landau gauge Gribov-Zwanziger action in the zero-temperature and zero-background limits. However, as it was pointed out in Ref.~\cite{Dudal:2017jfw}, it is not invariant under background gauge transformations. Not only is this at odds with the fact that both the gauge-fixing condition in the Landau-DeWitt gauge and the condition defining the corresponding first Gribov region are invariant under background gauge transformations, but it also prevents the implementation of center symmetry at finite temperature. Surprisingly, the one-loop background effective potential obtained in Ref.~\cite{Canfora:2015yia}  displays background gauge invariance but, as it was also clarified in Ref.~\cite{Dudal:2017jfw}, this is due to a missing term in the evaluation of the potential. 

To cure the lack of background gauge invariance, a new model action was also put forward in Ref.~\cite{Dudal:2017jfw}, based on a construction that preserves both BRST symmetry and background gauge invariance with the price however of introducing a Stueckelberg type field, not so easy to deal with, specially at finite temperature. In this article, we follow a sightly different route than that of Ref.~\cite{Dudal:2017jfw}. We first revisit the model of Ref.~\cite{Canfora:2015yia} and show how it can be very simply upgraded into a fully background gauge invariant one, that in addition correctly generates the one-loop results of that reference. This opens the way to the evaluation of higher order corrections in a manifestly background gauge invariant setting. We also try to discuss to which extent the model can be seen as a faithful implementation of the Gribov restriction for the Landau-DeWitt gauge.
 
 In Sec.~\ref{sec:model}, we introduce the model as a minimal, background gauge invariant modification of the action used in Ref.~\cite{Canfora:2015yia}. In Sec.~\ref{sec:1loop}, we compute the corresponding one-loop background effective potential for any gauge group and, in Sec.~\ref{sec:numerics}, we use it to investigate the deconfinement phase transition in SU(2) and SU(3) Yang-Mills theories. In particular, we study the impact on the transition temperatures of the use of color dependent Gribov parameters, as allowed by the model. Finally, in Sec.~\ref{sec:Gribov}, we provide a further motivation of the model by showing, at zero temperature and at leading order, how it is connected to the Gribov no-pole condition applied to the Landau-DeWitt gauge. We also discuss some of the difficulties that occur at finite temperature (similar to the ones discussed in Refs.~\cite{Cooper:2015sza,Cooper:2015bia} for the Landau gauge), when trying to interpret the model as arising from a faithful implementation of the Gribov restriction. More technical details are gathered in the Appendices. In particular, the various formulae needed for our analysis, including the case where certain Gribov parameters become negative, are given in Appendix~\ref{appsec:formulae}.

%%%%%
\section{A background gauge invariant Gribov-Zwanziger type model action}\label{sec:model}

We consider a pure Yang-Mills theory in $d$ Euclidean dimensions with a gauge group of dimension $d_G$. The Gribov-Zwanziger gauge-fixing procedure in the Landau gauge leads to the action
\beq\label{eq:GZ}
S & = & \int_x \bigg\{\frac{1}{4}F_{\mu\nu}^a F_{\mu\nu}^a+ih^a \partial_\mu A^a_\mu+\bar c^a \partial_\mu D^{ab}_\mu c^b\nonumber\\
& & \hspace{0.7cm}-\, \bar\omega^{ae}_\nu \partial_\mu D^{ab}_\mu \omega^{be}_\nu+\bar\varphi^{ae}_\nu \partial_\mu D^{ab}_\mu\varphi^{be}_\nu\nonumber\\
& & \hspace{0.7cm}-\,g \gamma^{1/2} f^{abc} A_\mu^a(\varphi^{bc}_\mu+\bar\varphi^{bc}_\mu)-\gamma dd_G\bigg\}\,,
\eeq
where $\smash{D^{ab}_\mu\equiv \partial_\mu\delta^{ab}-gf^{abc} A_\mu^c}$ denotes the covariant derivative in the adjoint representation. The first line of Eq.~(\ref{eq:GZ}) is nothing but the gauge-fixed action in the Landau gauge $\smash{\partial_\mu A^a_\mu=0}$ as it arises from the Faddeev-Popov procedure, while the second and third lines contain the corrections that arise from further restricting the functional integral to the first Gribov region, defined by the additional condition $-\partial_\mu D_\mu>0$.\footnote{For gauge field configurations satisfying the Landau gauge condition, the Faddeev-Popov operator $-\partial_\mu D_\mu$ is hermitian. Then, it makes sense to look for gauge field configurations such that this operator is, in addition, positive definite.} The complex conjugated bosonic fields $\varphi^{ab}_\nu$ and $\bar\varphi^{ab}_\nu$ together with the Grassmanian conjugated fields $\omega^{ab}_\nu$ and $\bar\omega^{ab}_\nu$ allow one to express this restriction in the form of a local field theory. Without loss of generality, they can be taken antisymmetric under exchange of their color indices. Finally, the parameter $\gamma$ is known as the Gribov parameter and is fixed using a saddle-point condition, see below.

%%%
\subsection{The problem}

In the background generalization of the Landau gauge, the so-called Landau-DeWitt gauge, one introduces a background gauge field configuration $\bar A_\mu^a$ and imposes the gauge-fixing condition 
\beq\label{eq:two}
\bar D^{ab}_\mu a^b_\mu=0\,,
\eeq 
where $\smash{a_\mu^a\equiv A_\mu^a-\bar A_\mu^a}$ is the fluctuation of the field $A_\mu^a$ about $\bar A_\mu^a$, and $\smash{\bar D^{ab}_\mu\equiv \partial_\mu\delta^{ab}-gf^{abc}\bar A_\mu^c}$ denotes the background covariant derivative. 

The corresponding Faddeev-Popov action can be obtained from the one in the Landau gauge using the simple mnemonic rule $\partial_\mu\to\bar D_\mu$ and $A_\mu^a\to a_\mu^a$. Based on this observation, the authors of Ref.~\cite{Canfora:2015yia} proposed the following action
\beq\label{eq:action_prop}
S & = & \int_x \bigg\{\frac{1}{4}F_{\mu\nu}^a F_{\mu\nu}^a+ih^a \bar D^{ab}_\mu a^b_\mu+\bar c^a \bar D_\mu^{ab} D^{bc}_\mu c^c\nonumber\\
& & \hspace{0.7cm}-\, (\omega^\dagger_\nu)^{ea} \bar D^{ab}_\mu D^{bc}_\mu \omega^{ce}_\nu+(\varphi^\dagger_\nu)^{ea} \bar D^{ab}_\mu D^{bc}_\mu\varphi^{ce}_\nu\nonumber\\
& & \hspace{0.7cm}-\,g \gamma^{1/2} f^{abc} a_\mu^a(\varphi^{bc}_\mu+\bar\varphi^{bc}_\mu)-\gamma dd_G\bigg\}\,,
\eeq
as a model action implementing the Gribov restriction in the case of the Landau-DeWitt gauge.\footnote{Here, as compared to Ref.~\cite{Canfora:2015yia}, we have considered a general group of dimension $d_G$, we have taken the gauge-fixing parameter to zero by introducing a Nakanishi-Lautrup field $h$ and we have redefined the Gribov parameter $\gamma$. We have also used a slightly different notation for the gauge field $A_\mu^a$ and the fluctuation $a_\mu^a$, more in line with the conventions of Ref.~\cite{Reinosa:2015gxn}.} For later convenience, we have used $\bar\varphi^{ae}_\nu=(\varphi^\dagger_\nu)^{ea}$ and $\bar\omega^{ae}_\nu=(\omega^\dagger_\nu)^{ea}$ to write the terms in the second line as color traces.

It was later realized in Ref.~\cite{Dudal:2017jfw} that the action (\ref{eq:action_prop}) cannot represent a faithful implementation of the restriction to the Gribov region in the Landau-DeWitt gauge. Indeed, despite the fact that the two conditions defining the Gribov region in this case, namely (\ref{eq:two}) and $\smash{-\bar D_\mu D_\mu>0}$, are invariant under the background gauge transformations
\beq
(\bar A^U_\mu)^a(x)t^a & = & U(x)\bar A_\mu^a(x)t^a U^\dagger(x)\!+\!\frac{i}{g}U(x)\partial_\mu U^\dagger(x),\label{eq:bgt1}\\
(a^U_\mu)^a(x)t^a & = & U(x) a^a_\mu(x) t^a U^\dagger(x),\label{eq:bgt2}
\eeq
the same does not hold for the action (\ref{eq:action_prop}). This can be seen as follows. In terms of coordinates, the adjoint transformation (\ref{eq:bgt2}) rewrites $(a^U_\mu)^a(x)={\cal U}_{ab}(x)a^b_\mu(x)$. To make the last line of Eq.~(\ref{eq:action_prop}) invariant under (\ref{eq:bgt2}), one should therefore require the field $\varphi^{ab}_\nu$ to transform as the product of two adjoint representations:
\beq\label{eq:tensort}
(\varphi^U_\nu)^{ab}(x) & = & {\cal U}_{ac}(x)\,{\cal U}_{bd}(x)\varphi^{cd}_\nu(x)\nonumber\\
& = & {\cal U}_{ac}(x)\varphi^{cd}_\nu(x)\,{\cal U}^\dagger_{db}(x)\,,
\eeq
where we used that the adjoint representation is real. In what follows, it will be convenient to use this transformation using a matrix notation, that is
\beq\label{eq:xit}
\varphi^U_\nu(x)={\cal U}(x)\varphi_\nu(x)\,{\cal U}^\dagger(x)\,.
\eeq
 The same transformation rule holds for $\bar\varphi_\nu$ since this field is the complex conjugate of $\varphi_\nu$ and the adjoint representation is real. Similarly, it is easily shown that
\beq
\bar D^U_\mu D^U_\mu={\cal U}(x)\bar D_\mu D_\mu\,{\cal U}^\dagger(x)\,.
\eeq
Therefore, the last term of the second line of Eq.~(\ref{eq:action_prop}) transforms as
\beq
& & {\rm tr}\,\Big((\varphi^U_\nu)^\dagger(x)\bar D^U_\mu D^U_\mu\varphi^U_\nu(x)\Big)\nonumber\\
& & \hspace{0.5cm}=\,{\rm tr}\,\Big({\cal U}(x)(\varphi_\nu)^\dagger(x)\bar D_\mu D_\mu[\varphi_\nu(x)\,{\cal U}^\dagger(x)]\Big).
\eeq
The ${\cal U}$-factors that originate from the left part of the transformation of $\varphi_\nu$ in Eq.~(\ref{eq:xit}) have cancelled out against those that appear when transforming the differential  operator $\bar D_\mu D_\mu$. In contrast, the ${\cal U}$-factors that originate from the right part of the transformation in Eq.~(\ref{eq:xit}) cannot be eliminated. Thus, the action (\ref{eq:action_prop}) is not invariant under the background gauge transformations (\ref{eq:bgt1})-(\ref{eq:bgt2}).\footnote{As already mentioned in the Introduction, the one-loop background effective potential obtained from the action (\ref{eq:action_prop}) in Ref.~\cite{Canfora:2015yia} appears nevertheless to be background gauge invariant. As was later observed in Ref.~\cite{Dudal:2017jfw}, this is due to the omission of some terms in the evaluation of the one-loop background effective potential that derives from the action (\ref{eq:action_prop}).}

To overcome these difficulties, a new action was put forward in Ref.~\cite{Dudal:2017jfw}, based on a BRST compatible model for the Gribov restriction, that automatically ensured the invariance under background gauge transformations. This construction is however not so easy to implement in practice because it requires the introduction of a SU(N)-valued field $h$ such that $A^h$ remains invariant under gauge transformations. This matrix valued field is usually handled by a Stueckelberg type field $\xi$ such that $h=e^{i\xi^at^a}$, which complicates the analysis. Moreover, at finite temperature, in order to preserve center symmetry, one needs a priori to integrate over fields $h$ that are periodic up to an element of the center of the gauge group, that is over topologically distinct sectors. How to achieve this in practice in terms of the Stuckelberg field is not completely clear. 

Here, a different route will be followed: we choose to sacrifice BRST symmetry with the benefit of obtaining a background gauge invariant setting that is easy to implement at finite temperature.\footnote{In the ideal scenario where one would select one Gribov copy per orbit, we expect BRST symmetry to be broken. We note however that, in the GZ scenario, a local BRST symmetry could be identified \cite{Capri:2016aqq}.}  We show that the action (\ref{eq:action_prop}) can be very simply upgraded into a background gauge invariant one and that the latter leads exactly to the same one-loop background effective potential as the one that was obtained in Ref.~\cite{Canfora:2015yia}. In fact our results will be slightly more general since our analysis will also reveal that it is possible to introduce color-dependent Gribov parameters without jeopardizing the background gauge invariance. We shall investigate this possibility in the application of the model to the deconfinement transition.

%%%
\subsection{A background gauge invariant model}\label{sec:IIB}
The problem discussed in the previous section could be summarized by saying that the breaking of background gauge invariance in the action (\ref{eq:action_prop}) stems from the fact that the operator $\bar D_\mu D_\mu$ is constructed out of covariant derivatives in the adjoint representation, whereas the objects this operator acts upon -- $\varphi_\nu$ and $\omega_\nu$ -- transform in a different representation, namely the tensor product of two adjoint representations. One possibility to restore background gauge invariance to the model action (\ref{eq:action_prop}) would be, therefore, to replace the operator $\bar D_\mu D_\mu$ by an operator $\bar{\cal D}_\mu {\cal D}_\mu$ where the covariant derivatives act now on the appropriate representation. With this approach, however, one would loose contact with the Faddeev-Popov operator $\bar D_\mu D_\mu$, which is at the heart of the definition of the first Gribov region. Moreover, in the Landau limit $\bar A\to 0$, one does not recover the usual Gribov-Zwanziger action.

Here, we shall restore background gauge-invariance using a different strategy that keeps contact with the Faddeev-Popov operator while recovering the well known $\bar A\to 0$ limit. The idea is to insert Wilson lines at appropriate places such that one of the two representations that enter the transformation of $\varphi_\nu$, more precisely the one acting to the right in Eq.~(\ref{eq:xit}), is not gauged. To this purpose, we replace the action (\ref{eq:action_prop}) by
\beq\label{eq:action_prop_mod}
S_{\rm new} & = & \int_x \bigg\{\frac{1}{4}F_{\mu\nu}^a F_{\mu\nu}^a+ih^a \bar D^{ab}_\mu a^b_\mu+\bar c^a \bar D_\mu^{ab} D^{bc}_\mu c^c\nonumber\\
& & \hspace{0.7cm}-\, (\hat\omega^\dagger_\nu)^{ea} \bar D^{ab}_\mu D^{bc}_\mu \hat\omega^{ce}_\nu+(\hat\varphi^\dagger_\nu)^{ea} \bar D^{ab}_\mu D^{bc}_\mu\hat\varphi^{ce}_\nu\nonumber\\
& & \hspace{0.7cm}-\,g \gamma^{1/2} f^{abc} a_\mu^a(\varphi^{bc}_\mu+\bar\varphi^{bc}_\mu)-\gamma dd_G\bigg\}\,,
\eeq
where we have introduced $\hat\varphi^{ac}_\nu(x)=\varphi^{ab}_\nu(x)L^{bc}_{\bar A,C}(x,x_0)$ and $\hat\omega^{ac}_\nu(x)=\omega^{ab}_\nu(x)L^{bc}_{\bar A,C}(x,x_0)$,\footnote{We redefine the field $\omega_\nu$ using the same Wilson line as that for $\varphi_\nu$ because these fields should be treated on an equal footing. Indeed, the r\^ole of the fields $\omega_\nu$ and $\bar\omega_\nu$ is to cancel a determinant generated by the integration over the fields $\varphi_\nu$ and $\bar\varphi_\nu$.} with
\beq\label{eq:Wilson}
L_{\bar A,C}(x,x_0)\equiv P\exp\left\{ig\int_C dy_\mu\,\bar A^a_\mu(y)\,T^a\right\}
\eeq
the Wilson line in the ajoint representation $t^a\mapsto T^a\equiv[t^a,\,\,]$ connecting the points $x_0$ to $x$ through the path $C$. It is easily checked that, under a background gauge transformation (\ref{eq:bgt1})-(\ref{eq:bgt2}), the Wilson line transforms as 
\beq
L_{\bar A^U\!\!,C}(x,x_0)={\cal U}(x)L_{\bar A,C}(x,x_0)\,{\cal U}^\dagger(x_0)\,,
\eeq 
and therefore
\beq
\hat\varphi^U_\nu(x)={\cal U}(x)\hat\varphi_\nu(x)\,{\cal U}^\dagger(x_0)\,.
\eeq
The crucial difference with Eq.~(\ref{eq:xit}) is that the right ${\cal U}$-factor of the transformation is $x$-independent. Consequently, one gets
\beq
& & {\rm tr}\,\Big((\hat\varphi^U_\nu)^\dagger(x)\bar D^U_\mu D^U_\mu\hat\varphi^U_\nu(x)\Big)\nonumber\\
& & \hspace{0.5cm}= {\rm tr}\,\Big({\cal U}(x_0)\hat\varphi^\dagger_\nu(x)\bar D_\mu D_\mu[\hat\varphi_\nu(x)\,{\cal U}^\dagger(x_0)]\Big).
\eeq
The remaining ${\cal U}$-factors are now $x$-independent and can be pulled out of the action of the covariant derivatives. They cancel owing to the cyclicity of the trace. Similar remarks apply to the term involving the fields $\omega_\nu$ and $\bar\omega_\nu$. This completes the proof of the background gauge invariance of the model action (\ref{eq:action_prop_mod}).

Before closing this section, we mention that there is a subtlety hidden in the previous discussion. Strictly speaking, if the background $\bar A_\mu^a$ is such that $\bar F_{\mu\nu}^a\neq 0$, objects such as the Wilson line or $\hat\varphi_\nu(x)$ and $\hat\omega_\nu(x)$ are not true functions of $x$ for they also depend on the chosen path $C$. In order to guarantee that our procedure makes sense, we should, therefore, specify what is meant by the action of the operator $\bar D_\mu D_\mu$ on this type of objects. We discuss this technical matter in Appendix \ref{app:Wilson}, where we also show that our construction is independent of the chosen path $C$ and in particular on the choice of $x_0$. The rest of the work will be concerned with constant backgrounds for which this subtlety does not appear.

%%%
\subsection{Choice of background and Cartan-Weyl basis}

The previous considerations apply a priori to any type of background, including instantonic backgrounds, provided the correct definitions are used (see Appendix A). However, for the finite temperature applications that we have in mind below, we shall restrict to backgrounds that explicitly preserve the space-time symmetries of the problem, namely Euclidean space-time translations and space rotations. Therefore, we assume that the background is temporal and constant over Euclidean space-time. In fact, without loss of generality, this type of backgrounds can be color-rotated to lie in the diagonal part of the algebra, the Cartan subalgebra:
\beq\label{eq:bg}
\beta g\bar A^a_\mu(x)t^a=\delta_{\mu0}\,r^j t^j\,,
\eeq
with $[t^j,t^{j'}]=0$. We have extracted a factor $\beta\equiv 1/T$ to make the components $r^j$ dimensionless.

For this type of backgrounds $\bar F_{\mu\nu}=0$ and the Wilson line becomes a true function of its endpoints, no longer depending on the chosen path in between. Choosing $x_0=0$, we arrive at
\beq
\hat\varphi_\nu(x)=\varphi_\nu(x)\,e^{i\frac{\tau}{\beta} r^j[t^j,\,\,]}\,.
\eeq
Similarly, the background covariant derivative rewrites
\beq
\bar D_\mu=\partial_\mu-iT\delta_{\mu0}\, r^j[t^j,\,\,]\,.
\eeq
These two quantities are the only sources for background dependence in the action (\ref{eq:action_prop_mod}). Since they involve only commutators with elements of the Cartan subalgebra, it is convenient to operate a change of basis from the usual Cartesian basis $it^a$ -- which we used to write the actions above -- to so-called Cartan-Weyl basis $it^\kappa$. 

By definition, the elements of a Cartan-Weyl basis diagonalize simultanously the adjoint action of the $t^j$'s
\beq
[t^j,t^\kappa]=\kappa_j t^\kappa\,.
\eeq
The color labels $\kappa$ should be seen as vectors in a space isomorphic to the Cartan subalgebra. They can take two types of values: either $\kappa=0^{(j)}$ is ``a zero'' in which case $t^\kappa$ is just different and convenient notation for $t^j$, or $\kappa=\alpha$ is a root of the algebra of the gauge group.\footnote{Below, we shall recall the zeros and the roots for the SU(2) and SU(3) groups. Note that there are as many zeros as there are dimensions in the Cartan subalgebra, hence the label $(j)$ to denote the various zeros.} The benefit of the Cartan-Weyl basis is that the background covariant derivative becomes diagonal, $\bar D_\mu^{\kappa\lambda}=\delta_{\kappa\lambda}\bar D_\mu^\kappa$, with
\beq
\bar D_\mu^\kappa=\partial_\mu -i T\delta_{\mu0}\,r^j\kappa^j\,.
\eeq
Similarly, the redefinition of the field $\varphi_\nu$ now appears as a simple multiplication by a phase factor depending on the rightmost color label of $\varphi_\nu$:
\beq\label{eq:mat}
\hat\varphi_\nu^{\kappa\xi}=\varphi^{\kappa\xi}_\nu\,e^{i\frac{\tau}{\beta} r^j\xi^j}\,.
\eeq
More details on how to change from the Cartesian basis to the Cartan-Weyl basis are given in Appendix~\ref{appsec:basis}. After some manipulations, we find
\beq\label{eq:action_prop_upgrade}
S_{\rm new} & = & \int_x\bigg\{\frac{1}{4}F_{\mu\nu}^{-\kappa} F_{\mu\nu}^\kappa+ih^{-\kappa} \bar D^{\kappa}_\mu a^\kappa_\mu-\,\bar D^{(-\kappa)}_\mu\bar c^{(-\kappa)} \bar D^{\kappa}_\mu c^\kappa\nonumber\\
& & \hspace{0.cm}+\,\bar D^{(-\kappa)}_\mu(\bar\omega^{(-\kappa)(-\xi)}_\nu e^{i\frac{\tau}{\beta}r\cdot\xi})\bar D^{\kappa}_\mu (\omega^{\kappa\xi}_\nu e^{-i\frac{\tau}{\beta}r\cdot\xi})\nonumber\\
& & \hspace{0.cm}-\,\bar D^{(-\kappa)}_\mu(\bar\varphi^{(-\kappa)(-\xi)}_\nu e^{i\frac{\tau}{\beta}r\cdot\xi}) \bar D^{\kappa}_\mu (\varphi^{\kappa\xi}_\nu e^{-i\frac{\tau}{\beta}r\cdot\xi})\nonumber\\
& & \hspace{0.cm}-\,ig f^{(-\kappa)\lambda\eta}\bar D^{(-\kappa)}_\mu\bar c^{(-\kappa)}a_\mu^\eta c^\lambda\nonumber\\
& & \hspace{0.cm}+\,igf^{(-\kappa)\lambda\eta}\bar D^{(-\kappa)}_\mu(\bar\omega^{(-\kappa)(-\xi)}_\nu e^{i\frac{\tau}{\beta}r\cdot\xi})\,a_\mu^\eta\omega^{\lambda\xi}_\nu e^{-i\frac{\tau}{\beta}r\cdot\xi}\nonumber\\
& & \hspace{0.cm}-\,igf^{(-\kappa)\lambda\eta}\bar D^{(-\kappa)}_\mu(\bar\varphi^{(-\kappa)(-\xi)}_\nu e^{i\frac{\tau}{\beta}r\cdot\xi})\,a_\mu^\eta\varphi^{\lambda\xi}_\nu e^{-i\frac{\tau}{\beta}r\cdot\xi}\nonumber\\
& & \hspace{0.cm}+\,ig \gamma^{1/2} f^{\kappa\lambda\eta} a_\mu^{\kappa}(\varphi^{\lambda\eta}_\mu+\bar\varphi^{\lambda\eta}_\mu)-\gamma dd_G\bigg\},
\eeq
with $F_{\mu\nu}^\kappa=\partial_\mu A_\nu^\kappa-\partial_\nu A_\mu^\kappa -ig  f^{(-\kappa)\lambda\eta}A_\mu^\lambda A_\nu^\eta$ and $[t^\lambda,t^\eta]=f^{(-\kappa)\lambda\eta}t^\kappa$. So defined, the structure constants are antisymmetric and conserve color in the following sense: $f^{\kappa\lambda\tau}=0$ if $\kappa+\lambda+\tau\neq 0$ \cite{Reinosa:2015gxn}.

Of course, since we have restricted to backgrounds of the form (\ref{eq:bg}), we should restrict to transformations that preserve this form. Those read
\beq
{r'}^{j}=r^j+\bar\alpha^j\,,\label{eq:transfo1}
\eeq
together with
\beq
{X'}^\kappa(x) & = & e^{i\frac{\tau}{\beta}\bar\alpha\cdot\kappa}X^\kappa(x)\,,\label{eq:transfo2}\\
{X'}^{\kappa\lambda}(x) & = & e^{i\frac{\tau}{\beta}\bar\alpha\cdot(\kappa+\lambda)}X^{\kappa\lambda}(x)\,.\label{eq:transfo3}
\eeq
The $\bar\alpha$'s are certain vectors that we do not need to specify further here, see for instance Ref.~\cite{Reinosa:2015gxn} for more details. Using the property $\smash{{\bar D}_\mu^{'\kappa} {X'}^\kappa(x)=e^{i\frac{\tau}{\beta}\bar\alpha\cdot\kappa}{\bar D}_\mu^{\kappa} {X}^\kappa(x)}$ and the fact that $f^{\kappa\lambda\eta}$ conserves color, one easily checks that the action (\ref{eq:action_prop_upgrade}) is invariant under the background gauge transformations (\ref{eq:transfo1})-(\ref{eq:transfo3}). Again, the r\^ole of the phase factors originating from the Wilson lines is crucial. The action can be equivalently rewritten as
\beq\label{eq:action_prop_upgrade_2}
S_{\rm new} & = & \int_x\bigg\{\frac{1}{4}F_{\mu\nu}^{-\kappa} F_{\mu\nu}^\kappa+ih^{-\kappa} \bar D^{\kappa}_\mu a^\kappa_\mu-\,\bar D^{(-\kappa)}_\mu\bar c^{(-\kappa)} \bar D^{\kappa}_\mu c^\kappa\nonumber\\
& & \hspace{0.7cm}+\,\bar D^{(-\kappa-\xi)}_\mu \bar\omega^{(-\kappa)(-\xi)}_\nu \bar D^{\kappa+\xi}_\mu \omega^{\kappa\xi}_\nu\nonumber\\
& & \hspace{0.7cm}-\,\bar D^{(-\kappa-\xi)}_\mu \bar\varphi^{(-\kappa)(-\xi)}_\nu \bar D^{\kappa+\xi}_\mu \varphi^{\kappa\xi}_\nu\nonumber\\
& & \hspace{0.7cm}-\,ig f^{(-\kappa)\lambda\eta}\bar D^{(-\kappa)}_\mu\bar c^{(-\kappa)}a_\mu^\eta c^\lambda\nonumber\\
& & \hspace{0.7cm}+\,igf^{(-\kappa)\lambda\eta}\bar D^{(-\kappa-\xi)}_\mu \bar\omega^{(-\kappa)(-\xi)}_\nu  a_\mu^\eta\omega^{\lambda\xi}_\nu\nonumber\\
& & \hspace{0.7cm}-\,igf^{(-\kappa)\lambda\eta}\bar D^{(-\kappa-\xi)}_\mu \bar\varphi^{(-\kappa)(-\xi)}_\nu  a_\mu^\eta\varphi^{\lambda\xi}_\nu\nonumber\\
& & \hspace{0.7cm}+\,ig \gamma^{1/2} f^{\kappa\lambda\eta} a_\mu^{\kappa}(\varphi^{\lambda\eta}_\mu+\bar\varphi^{\lambda\eta}_\mu)-\gamma dd_G\bigg\},
\eeq
which makes the invariance even more explicit.

In what follows, we take the action (\ref{eq:action_prop_upgrade_2}) as our model for a Gribov-Zwanziger type model action invariant under background gauge transformations.  In Sec.~\ref{sec:Gribov}, we provide a further motivation for the model by showing that, at zero-temperature and to one-loop accuracy, it is related to the Gribov no-pole condition applied to the Landau-DeWitt gauge.\footnote{We should mention, however, that it is far from obvious that our proposal or the one in Ref.~\cite{Dudal:2017jfw} correspond to faithful implementations of the Gribov restriction at finite temperature. We briefly discuss this issue in Sec.~\ref{sec:Gribov}.}

Moreover, in sec. \ref{sec:mappin_Landau}, we show that, for vanishing temperatures, the configuration-space correlation functions of the model 
\eqref{eq:action_prop_upgrade_2} are related to those associated to the Gribov-Zwanziger action in the Landau gauge, implying that addition of a 
background field does not spoil renormalizability at $T=0$. As one moves to finite temperature, one should also expect renormalizability to hold, for 
the thermal contributions always come with a statistical factor, which works as a smooth UV cutoff.

%%%
\subsection{Color-dependent Gribov parameters}

Before closing this section, it should be mentioned that the model can, and will, be extended by introducing color-dependent Gribov parameters $\gamma_\kappa$ without affecting the background gauge invariance (\ref{eq:transfo1})-(\ref{eq:transfo3}):\footnote{The reason why the color label of $\gamma$ is the one associated to $a_\mu$ is that the fields $\varphi_\mu$ are just auxiliary fields that help localizing the action.}
\beq\label{eq:action_prop_upgrade_3}
S_{\rm new} & = & \int_x\bigg\{\frac{1}{4}F_{\mu\nu}^{-\kappa} F_{\mu\nu}^\kappa+ih^{-\kappa} \bar D^{\kappa}_\mu a^\kappa-\,\bar D^{(-\kappa)}_\mu\bar c^{(-\kappa)} \bar D^{\kappa}_\mu c^\kappa\nonumber\\
& & \hspace{0.5cm}+\,\bar D^{(-\kappa-\xi)}_\mu \bar\omega^{(-\kappa)(-\xi)}_\nu \bar D^{\kappa+\xi}_\mu \omega^{\kappa\xi}_\nu\nonumber\\
& & \hspace{0.5cm}-\,\bar D^{(-\kappa-\xi)}_\mu \bar\varphi^{(-\kappa)(-\xi)}_\nu \bar D^{\kappa+\xi}_\mu \varphi^{\kappa\xi}_\nu\nonumber\\
& & \hspace{0.5cm}-\,ig f^{(-\kappa)\lambda\eta}\bar D^{(-\kappa)}_\mu\bar c^{(-\kappa)}a_\mu^\eta c^\lambda\nonumber\\
& & \hspace{0.5cm}+\,igf^{(-\kappa)\lambda\eta}\bar D^{(-\kappa-\xi)}_\mu \bar\omega^{(-\kappa)(-\xi)}_\nu  a_\mu^\eta\omega^{\lambda\xi}_\nu\nonumber\\
& & \hspace{0.5cm}-\,igf^{(-\kappa)\lambda\eta}\bar D^{(-\kappa-\xi)}_\mu \bar\varphi^{(-\kappa)(-\xi)}_\nu  a_\mu^\eta\varphi^{\lambda\xi}_\nu\nonumber\\
& & \hspace{0.5cm}+\,ig \gamma^{1/2}_\kappa f^{\kappa\lambda\eta} a_\mu^{\kappa}(\varphi^{\lambda\eta}_\mu+\bar\varphi^{\lambda\eta}_\mu)-d\sum_\kappa\gamma_\kappa\bigg\}.
\eeq
We will see below that the Gribov parameters are all degenerate at zero temperature. At finite temperature, in contrast, there is no reason for them to remain equal and, therefore, it will be interesting to compare the situation where a unique Gribov parameter is attributed to all color modes with the one where Gribov parameters are allowed to depend on color.

Our main focus being the study of the deconfinement transition it is however  of crucial importance to preserve the invariance under so-called Weyl transformations,\footnote{These are finite color rotations that leave the Cartan subalgebra globally invariant.} because only then the background field, as obtained from the minimization of the background effective potential is an order parameter for center symmetry \cite{Reinosa:2015gxn,Herbst:2015ona}. Since the Weyl transformations typically connect certain roots $\alpha$ and $\beta$ with each other, a simple way to ensure Weyl symmetry is to impose that $\smash{\gamma_\alpha=\gamma_\beta}$ for such roots. If one also wants to preserve invariance under charge conjugation, one possibility is to impose that $\smash{\gamma_\alpha=\gamma_{-\alpha}}$. In what follows, we shall consider groups where Weyl transformations and charge conjugation allow to connect all roots with each other and therefore we introduce a single Gribov parameter $\gamma_{\rm ch}$ for all these ``charged'' modes. In contrast for each ``neutral'' mode,\footnote{The terminology ``charged'' and ``neutral'' arises from the fact that $\kappa\cdot r T$ can be seen as a color-dependent imaginary chemical potential.} corresponding to $\smash{\kappa=0^{(j)}}$, we can a priori introduce a different Gribov parameter $\gamma_{0^{(j)}}$.

In fact, this choice of Gribov parameters is just a sufficient condition to ensure Weyl symmetry but it is not necessary. Weyl symmetry is more generally preserved in the following sense: the action (\ref{eq:action_prop_upgrade_3}) is invariant under a Weyl transformation that exchanges $\alpha$ and $\beta$ provided one also performs the transformation $\gamma_\alpha\leftrightarrow\gamma_\beta$. This symmetry is trivially inherited by the background effective potential due to the extremization needed to determine the Gribov parameters, which are then promoted to functions of the background. I.e, when action \eqref{eq:action_prop_upgrade_3} is evaluated for the values of the $\gamma$'s obtained through this process, Weyl invariance is guaranteed in the usual sense. The same remarks apply to charge conjugation.\\

In summary, we shall study three different scenarios, all compatible with background gauge invariance, including Weyl invariance:\\

\hglue4mm {\bf Degenerate case:} all $\gamma_\kappa$'s taken equal.\\

\hglue4mm {\bf Partially degenerate:} all $\gamma_\alpha$'s taken equal.\\

\hglue4mm {\bf Non-degenerate case:} all $\gamma_\kappa$'s taken different.\\

\noindent{For simplicity, we shall however assume that $\gamma_\kappa=\gamma_{-\kappa}$, even in the third scenario.

%%%%%
\section{The model at one-loop}\label{sec:1loop}
In this section, we evaluate the background effective potential and the corresponding gap equation(s) at one-loop order, for any gauge group.

%%%
\subsection{Background effective potential}
The field $a_\mu^\kappa(x)$ contains both real ($a_\mu^{0^{(j)}}$) and complex conjugated components ($a_\mu^\alpha(x)$ and  $a_\mu^{-\alpha}(x)$). Moreover $\varphi^{\eta\xi}_\rho(x)$ and $\bar\varphi^{(-\eta)(-\xi)}_\rho(x)$ are also complex conjugate of each other, see Appendix~\ref{appsec:basis}. Following Appendix~\ref{appsec:gaussian}, one way to deal with the presence of both real and complex conjugated degrees of freedom is to write the quadratic part of the action in the bosonic sector as
\beq\label{eq:M0}
\frac{1}{2}\int_{x,y} \!\!\chi^\dagger(x)\,{\cal M}(x-y)\,\chi(y)\,,
\eeq
with $\chi^\dagger(x)=(a^\kappa_\mu(x),h^\lambda(x),\varphi^{\eta\xi}_\rho(x),\bar\varphi^{(-\eta)(-\xi)}_{\rho}(x))^*$, while in the Grassmannian sector, it is enough to write
\beq
\int_{x,y} \!\!{\bar \Upsilon}^{\rm t}(x)\,{\cal N}(x-y)\Upsilon(y)\,,
\eeq
with $\Upsilon^{\rm t}(x)=(c^\kappa(x),\omega^{\eta\xi}_\rho (x))$. 

The one-loop background effective potential reads
\beq
V(\bar A,\{\gamma_\kappa\})\!=\!-d\sum_\kappa\gamma_\kappa+\frac{1}{2}\ln{\rm det}\,{\cal M}-\ln{\rm det}\,{\cal N}.
\eeq
\begin{widetext}
In Fourier space, Eq.~(\ref{eq:M0}) rewrites $\frac{1}{2}\int_Q (\chi(Q))^\dagger {\cal M}(Q)\chi(Q)$ with $(\chi(Q))^\dagger=(a^\kappa_\mu(Q),h^\lambda(Q),\varphi^{\eta\xi}_\rho(Q),\bar\varphi^{(-\eta)(-\xi)}_{\rho}(Q))^*$ and
\beq\label{eq:M}
{\cal M}(Q)=\left(\begin{array}{cccc}
Q^2_\kappa P^\perp_{\mu\mu'}(Q^\kappa)\delta_{\kappa\kappa'} & -Q_{\mu}^{\kappa}\delta_{\kappa\lambda'} & ig\gamma^{1/2}_{\kappa} f^{(-\kappa)\eta'\xi'}\delta_{\mu\rho'} & -ig\gamma^{1/2}_{\kappa} f^{\kappa\eta'\xi'}_*\delta_{\mu\bar\rho'}\\
Q_{\mu'}^{\kappa'}\delta_{\kappa'\lambda} & 0 & 0 & 0 \\
-ig\gamma^{1/2}_{\kappa'} f^{(-\kappa')\eta\xi}_*\delta_{\mu'\rho} & 0 & -Q^2_{\eta'+\xi'}\delta_{\eta\eta'}\delta_{\xi\xi'}\delta_{\rho\rho'} & 0\\
ig\gamma^{1/2}_{\kappa'} f^{\kappa'\eta\xi}\delta_{\mu'\bar\rho} & 0 & 0 & -Q^2_{-\eta'-\xi'}\delta_{\eta\eta'}\delta_{\xi\xi'}\delta_{\rho\rho'}
\end{array}\right),
\eeq
where we have introduced the shifted momenta $Q^\kappa_\mu\equiv Q_\mu+r\cdot\kappa\,T\delta_{\mu0}$ and we have used $f^{(-\kappa)(-\eta)(-\xi)}=-f^{\kappa\eta\xi}_*$, where the subscript $*$ on $f^{\kappa\lambda\tau}_*$ denotes complex conjugation. In order to compute the determinant of $M$, we consider it as a block matrix of the form $(A\,\,B\,|\,C\,\,D)$, with $A$ and $D$ invertible, and use
\beq\label{eq:schur}
{\rm det}\,{\cal M}={\rm det}\,D \times {\rm det}\,(A-BD^{-1}C)={\rm det}\,A \times {\rm det}\,(D-CA^{-1}B)\,.
\eeq
A simple calculation shows that
\beq
A-BD^{-1}C=\left(\begin{array}{cc}
Q^2_\kappa P^\perp_{\mu\mu'}(Q^\kappa)\delta_{\kappa\kappa'}+2g^2\gamma^{1/2}_{\kappa}\gamma^{1/2}_{\kappa'} f^{(-\kappa)\eta\xi}f^{(-\kappa')\eta\xi}_*Q^{-2}_{\eta+\xi}\delta_{\mu\mu'}  & -Q_{\mu}^{\kappa}\delta_{\kappa\lambda'}\\
Q_{\mu'}^{\kappa'}\delta_{\kappa'\lambda} & 0
\end{array}\right)\!,
\eeq
where a summation over $\eta$ and $\xi$ is implied in the first element. We next use that that the structure constants conserve color, to write $\gamma^{1/2}_{\kappa}\gamma^{1/2}_{\kappa'}f^{(-\kappa)\eta\xi}f^{(-\kappa')\eta\xi}_*Q^{-2}_{\eta+\xi}=\gamma^{1/2}_{\kappa}\gamma^{1/2}_{\kappa'}f^{(-\kappa)\eta\xi}f^{(-\kappa')\eta\xi}_*Q^{-2}_\kappa=C_{\rm ad}\gamma_{\kappa}\delta_{\kappa\kappa'}Q^{-2}_\kappa$, where $C_{\rm ad}$ denotes the Casimir of the adjoint representation. We obtain
\beq
A-BD^{-1}C=\left(\begin{array}{cc}
\big[m^4_\kappa Q_\kappa^{-2}P^\parallel_{\mu\mu'}(Q_\kappa)+(Q^2_\kappa+m^4_\kappa Q_\kappa^{-2})P^\perp_{\mu\mu'}(Q^\kappa)\big]\delta_{\kappa\kappa'} & -Q_{\mu}^{\kappa}\delta_{\kappa\lambda'}\\
Q_{\mu'}^{\kappa'}\delta_{\kappa'\lambda} & 0 \\
\end{array}
\right)\!,
\eeq
where we have defined $m^4_\kappa\equiv 2g^2C_{\rm ad} \gamma_{\kappa}$. Using the second form of Eq.~(\ref{eq:schur}), we find
\beq
{\rm det}\,(A-BD^{-1}C)=\prod_\kappa \frac{m^4_\kappa}{Q_\kappa^2}\left(\frac{Q_\kappa^4+m^4_\kappa}{Q_\kappa^2}\right)^{d-1}\frac{Q_\kappa^4}{m^4_\kappa}=\prod_\kappa \left(\frac{Q_\kappa^4+m^4_\kappa}{Q_\kappa^2}\right)^{d-1}Q_\kappa^2\,,
\eeq
and then
\end{widetext}
\beq
{\rm det}\,{\cal M}={\rm det}\,D\times \prod_\kappa \left(\frac{Q_\kappa^4+m^4_\kappa}{Q_\kappa^2}\right)^{d-1}Q_\kappa^2\,.
\eeq
On the other hand, it is trivially shown that
\beq
{\rm det}\,{\cal N}=({\rm det}\,D)^{1/2}\times \prod_\kappa Q_\kappa^2\,.
\eeq
Therefore
\beq\label{eq:V}
V(r,\{m_\kappa^4\})=-\frac{d}{2}\frac{\sum_\kappa m^4_\kappa}{g^2C_{\rm ad}} & + & \frac{d-1}{2}\sum_\kappa \int_Q^T \ln \frac{Q_\kappa^4+m^4_\kappa}{Q_\kappa^2}\nonumber\\
& - & \frac{1}{2}\sum_\kappa \int_Q^T \ln Q_\kappa^2\,,
\eeq
where we have introduced the notations
\beq
\int_Q^T \equiv \mu^{2\epsilon} T\sum_n \int_q  \quad {\rm and} \quad \int_q \equiv \int\frac{d^{d-1}q}{(2\pi)^{d-1}}\,,
\eeq
with $d=4-2\epsilon$.

In the SU(2) case, and assuming that all the Gribov parameters $\smash{\gamma_\kappa\propto m^4_\kappa}$ are equal, the expression in Eq. \eqref{eq:V} is exactly the one-loop potential obtained in Ref.~\cite{Canfora:2015yia} but this time obtained from action (\ref{eq:action_prop_upgrade_3}) and not from (\ref{eq:action_prop}). So it seems that the terms missed in the computation of the one-loop effective potential from action \eqref{eq:action_prop}, as performed in Ref.~\cite{Canfora:2015yia}, are exactly eaten up by the extra phase factors introduced in Eq.~\eqref{eq:action_prop_upgrade}. Besides providing a proper justification to the one-loop formula of Ref.~\cite{Canfora:2015yia}, our model opens the way to the evaluation of higher order corrections in a background gauge invariant setting, which we plan to investigate in a future work.

%%%
\subsection{Gribov parameters}
The Gribov parameters are usually obtained from a saddle-point approximation, which boils down to extremizing the potential, not only with respect to the background but also with respect to the Gribov parameters themselves. It is important to realize that, even though the Gribov parameters will, at least in the present setting, always be found real, some of them could -- and will -- become negative. For the various cases studied, we find the following gap equations:\\

{\bf Degenerate case:}
\beq
0=\sum_\kappa \left[\frac{d}{d-1}\frac{1}{g^2C_{\rm ad}}-\hat J_\kappa(m^4)\right],
\eeq

{\bf Partially degenerate case:}
\beq
& & \forall j,\,0=\frac{d}{d-1}\frac{1}{g^2C_{\rm ad}}-\hat J_{0^{(j)}}(m^4_{0^{(j)}})\,,\nonumber\\
& & 0=\sum_\alpha\left[\frac{d}{d-1}\frac{1}{g^2C_{\rm ad}}-\hat J_\alpha(m^4_{\rm Ch})\right],
\eeq

{\bf Non-degenerate case:} 
\beq
& & \forall j,\,0=\frac{d}{d-1}\frac{1}{g^2C_{\rm ad}}-\hat J_{0^{(j)}}(m^4_{0^{(j)}})\,,\nonumber\\
& & \forall \alpha,\,0=\frac{d}{d-1}\frac{2}{g^2C_{\rm ad}}-(\hat J_\alpha(m^4_\alpha)+\hat J_{-\alpha}(m^4_\alpha))\,,
\eeq
where we have introduced the sum-integral
\beq
\hat J_\kappa(m^4)\equiv\int_Q^T\frac{1}{Q^4_\kappa+m^4}\,.
\eeq
Note that, because $\hat J_\kappa(m^4)=\hat J_{-\kappa}(m^4)$, the second equation in the non-degenerate case simplifies to
\beq
& & \forall \alpha,\,0=\frac{d}{d-1}\frac{1}{g^2C_{\rm ad}}-\hat J_\alpha(m^4_\alpha)\,.
\eeq

\subsubsection{Zero temperature limit}
In the zero-temperature limit, because the shifted momentum $Q_\kappa$ can always be shifted back to $Q$ via a change of variables, all Gribov parameters obey the same equation
\beq\label{eq:gap0}
0=\frac{d}{d-1}\frac{1}{g^2C_{\rm ad}}-\hat J(m^4_{\rm vac})\,,
\eeq
with
\beq
\hat J(m^4)\equiv\int_Q\frac{1}{Q^4+m^4}
\eeq
the zero-temperature version of $\hat J_\kappa(m^4)$. This integral, if restricted to real Gribov parameters, is defined only for $m^4>0$; its evaluation is recalled in Appendix~\ref{appsec:formulae}. We then arrive at the well known zero-temperature gap equation \cite{Vandersickel:2011zc}
\beq\label{eq:32}
0=1-\frac{3g^2C_{\rm ad}}{64\pi^2}\left[\frac{1}{\epsilon}+\frac{1}{2}\ln\frac{\bar\mu^4}{m^4_{\rm vac}}+\frac{5}{6}\right],
\eeq
which can be renormalized by setting (minimal subtraction scheme)
\beq\label{eq:renorm}
\frac{1}{g^2C_{\rm ad}}=\frac{1}{g^2(\bar\mu)C_{\rm ad}}+\frac{3}{64\pi^2}\frac{1}{\epsilon}\,.
\eeq
The renormalized equation reads
\beq
0=1-\frac{3g^2(\bar\mu)C_{\rm ad}}{128\pi^2}\left[\ln\frac{\bar\mu^4}{m^4_{\rm vac}}+\frac{5}{3}\right],
\eeq
and is solved as
\beq\label{eq:mass}
m^4_{\rm vac}=\bar\mu^4\exp\left(\frac{5}{3}-\frac{128\pi^2}{3g^2(\bar\mu)C_{\rm ad}}\right).
\eeq
From Eq.~(\ref{eq:renorm}), we find that the renormalized coupling runs with the beta function
\beq
\beta_{g^2}\equiv\bar\mu\frac{dg^2}{d\bar\mu}=-g^4\bar\mu\frac{d(1/g^2)}{d\bar\mu}=-\frac{3g^4C_{\rm ad}}{32\pi^2}\,.
\eeq
The sign is compatible with asymptotic freedom but the coefficient is not the expected one at order $g^4$. This happens due to other $g^4$ contributions arising from the two-loop corrections to the background effective potential. The two-loop gap equation has been determined and renormalized at zero temperature in Ref.~\cite{Gracey:2005cx}. At this order the Gribov parameter is also renormalized. We expect the same renormalization factors to renormalize the finite temperature two-loop gap equation. We shall consider this equation in a subsequent work together with the two-loop corrections to the background effective potential.

 In principle we could use Eq.~(\ref{eq:mass}) to fix the scale $m_{\rm vac}$ in terms of the known value of $g(\bar \mu)$ in the minimal subtraction scheme at some large ultraviolet scale $\bar\mu=\bar\mu_0$. However, since the running of $g(\bar\mu)$ does not coincide, not even at order $g^4$, with the true running, we expect large errors in the scale setting. We therefore postpone this question to a forthcoming two-loop study -- where the running coupling should be exact at leading order. In what follows,  we express all our results in units of $m_{\rm vac}$. This also allows for an easy comparison with Ref.~\cite{Canfora:2015yia}. We finally mention that the solution $m^4_{\rm vac}$ is unique, given the renormalized coupling at the scale $\bar\mu$. This means that not only the Gribov parameters all obey the same equation at zero-temperature but also that they all become equal, as announced above.

\subsubsection{Finite temperature case}

Following Ref.~\cite{Canfora:2015yia}, we can always parametrize the gap equations at finite temperature in terms of the solution $m^4_{\rm vac}$ at zero temperature. Subtracting the zero temperature equations from the finite temperature ones we find the following gap equations\\

{\bf Degenerate case:}
\beq
0=\sum_\kappa \Delta \hat J_\kappa(m^4;m_{\rm vac}^4)\,,
\eeq

{\bf Partially degenerate:} 
\beq
& & \forall j,\, 0=\Delta \hat J_{0^{(j)}}(m_{0^{(j)}}^4;m_{\rm vac}^4)\,,\nonumber\\
 & & 0=\sum_\alpha \Delta \hat J_\alpha(m_{\rm Ch}^4;m_{\rm vac}^4)\,,
 \eeq
 
{\bf Non-degenerate case:}
\beq
& & \forall j,\,0=\forall j,\, \Delta \hat J_{0^{(j)}}(m_{0^{(j)}}^4;m_{\rm vac}^4)\,,\nonumber\\
& & \forall \alpha,\, 0=\Delta \hat J_\alpha(m_\alpha^4;m_{\rm vac}^4)\,,
\eeq
where we have introduced the UV finite difference
\beq
& &\Delta \hat J_\kappa(m^4;m_{\rm vac}^4)\equiv\hat J_\kappa(m^4)-\hat J_{T=0}(m^4_{\rm vac})\nonumber\\
& & \hspace{0.2cm}=\,\int_Q^T \frac{1}{((\omega_n+Tr\cdot\kappa)^2+q^2)^2+m^4}-\int_Q \frac{1}{Q^4+m^4_{\rm vac}}\,.\nonumber\\
\eeq
Some useful remarks are in order here. First of all, $\Delta \hat J_{0^{(j)}}(m^4;m_{\rm vac}^4)$ is a strictly decreasing function over the interval $\smash{m^4\,\in\, \left]0,+\infty\right[}$ that diverges positively as $\smash{m^4\to 0^+}$ and becomes negative as $\smash{m^4\to +\infty}$. This implies that the gap equation for the neutral Gribov parameter $m^4_{0^{(j)}}$ has a unique solution and therefore that all neutral Gribov parameters coincide. We shall denote their common value $m^4_{\rm N}$ in the following. The same behavior holds for the function $\sum_\kappa \Delta \hat J_\kappa(m^4;m_{\rm vac}^4)$ and then the gap equation for the degenerate Gribov parameter $m^4$ has a unique solution. It also follows that $m^4_{\rm N}$ and $m^4$ are strictly positive. 

Similar conclusions hold for $m^4_\alpha$ and $m^4_{\rm Ch}$ with the noticeable difference that these parameters can become negative. Indeed, it is easily checked that $\Delta \hat J_\alpha(m^4;m_{\rm vac}^4)$ is a strictly decreasing function over the interval $\smash{m^4\,\in\, \left]-M^4_{r\cdot\alpha},+\infty\right[}$, with $\smash{M_{r\cdot\alpha}^4\equiv {\rm min}_{n\in\mathds{Z}}(2\pi n+r\cdot\alpha)^4T^4}$.  It diverges positively as $\smash{m^4\to -M_{r\cdot\alpha}^{4+}}$ and becomes negative as $\smash{m^4\to\infty}$. From this it follows that the gap equation for $m^4_\alpha$ has a unique solution for given values of the temperature and the background but this solution can become negative since the only constraint is that it should remain strictly larger than $-M_{r\cdot\kappa}^4$. In fact, we can determine at which temperature $m^4_\alpha$ may vanish. We just need to enforce a zero solution in the corresponding equation, namely 
\beq\label{eq:T0}
0=\Delta \hat J_\alpha(0;m_{\rm vac}^4)\,.
\eeq 
Similar considerations apply to $m^2_{\rm Ch}$ but now the function $\sum_\alpha \hat J_\alpha(m^4;m_{\rm vac}^4)$ diverges as $m^4\to -{\rm min}_{\alpha}M_{r\cdot\alpha}^4$. Again the temperature at which $m^4_{\rm Ch}$ may vanish can be obtained by solving the equation 
\beq
0=\sum_\alpha \Delta \hat J_\alpha(0;m_{\rm vac}^4)\,.
\eeq

In practice, when evaluating $\Delta \hat J_\kappa(m^4;m_{\rm vac}^4)$, we need to distinguish the case where $m^4>0$ and the case where $ -M^4_{r\cdot\kappa}<m^4<0$. These two cases are discussed in Appendix~\ref{appsec:formulae}. For $m^4>0$, we find
\beq\label{eq:toto4}
& & \Delta \hat J_\kappa(m^4;m^4_{\rm vac})=\frac{1}{32\pi^2}\ln\frac{m^4_{\rm vac}}{m^4}\nonumber\\
& & +\frac{1}{4im^2}\!\int_q \frac{1}{\sqrt{q^2+im^2}}\frac{\cos(r\cdot\kappa)-e^{-\beta\sqrt{q^2+im^2}}}{\cos(r\cdot\kappa)-\cosh(\beta\sqrt{q^2+im^2})}\nonumber\\
& & -\frac{1}{4im^2}\!\int_q\frac{1}{\sqrt{q^2-im^2}}\frac{\cos(r\cdot\kappa)-e^{-\beta\sqrt{q^2-im^2}}}{\cos(r\cdot\kappa)-\cosh(\beta\sqrt{q^2-im^2})}\,.\nonumber\\
\eeq
For $ -M^4_{r\cdot\kappa}<m^4\equiv -M^4<0$, we find instead
\beq
& & \Delta \hat J_\kappa(m^4;m^4_{\rm vac})=\frac{1}{32\pi^2}\ln\frac{m^4_{\rm vac}}{M^4}\nonumber\\
& & -\frac{1}{4M^2}\!\!\int_{q<M}\!\!\!\frac{1}{\sqrt{M^2-q^2}}\frac{\sin(\beta\sqrt{M^2-q^2})}{\cos(r\cdot\kappa)-\cos(\beta\sqrt{M^2-q^2})}\nonumber\\
& & -\frac{1}{4M^2}\!\!\int_{q>M}\!\!\!\frac{1}{\sqrt{q^2-M^2}}\frac{\cos(r\cdot\kappa)-e^{-\beta\sqrt{q^2-M^2}}}{\cos(r\cdot\kappa)-\cosh(\beta\sqrt{q^2-M^2})}\nonumber\\
& & +\frac{1}{4M^2}\!\!\int_q \frac{1}{\sqrt{q^2+M^2}}\frac{\cos(r\cdot\kappa)-e^{-\beta\sqrt{q^2+M^2}}}{\cos(r\cdot\kappa)-\cosh(\beta\sqrt{q^2+M^2})}\,.\nonumber\\
\eeq
For practical purposes, it is convenient to absorb the integrable singularity (in the second integral) as $q\to M$ using the change of variables $u=\sqrt{q^2-M^2}$. For consistency, we apply similar changes of variables to the other two integrals.

%%%
\subsection{Finite form of the effective potential}

Finally, the integrals that enter the one-loop potential are also well known and recalled in Appendix \ref{appsec:formulae}. Using Eq~(\ref{eq:gap0}), we find $V(r,\{m^4_\kappa\})=\sum_\kappa V_\kappa(r,m^4_\kappa)$ with
\beq
V_\kappa(r,m^4)=\frac{d-1}{2}\Delta\hat K_\kappa(m^4,m^4_{\rm vac})-\frac{d}{4} \Delta\hat K_\kappa(0,m^4_{\rm vac})\nonumber\\
\eeq
and
\beq
\Delta \hat K^\kappa(m^4,m^4_{\rm vac})\equiv \int_Q^T\ln (Q_\kappa^4+m^4)- \int_Q \frac{m^4}{Q^4+m^4_{\rm vac}}\,.\nonumber\\
\eeq
It is easily checked that this expression is UV finite, up to a quartic divergence that vanishes in dimensional regularization. More precisely, in the case where $m^4$ is positive, we find (see Appendix~\ref{appsec:formulae})
\beq
& & \Delta\hat K_\kappa(m^4;m^4_{\rm vac})=\frac{m^4}{32\pi^2}\left[\ln\frac{m^4_{\rm vac}}{m^4}+1\right]\nonumber\\
& & \hspace{0.5cm}+\,T\int_q\,\ln(e^{-2\beta\sqrt{q^2+im^2}}\nonumber\\
& & \hspace{2.0cm}-\,2e^{-\beta\sqrt{q^2+im^2}}\cos(r\cdot\kappa)+1)\nonumber\\
& & \hspace{0.5cm}+\,T\int_q\,\ln(e^{-2\beta\sqrt{q^2-im^2}}\nonumber\\
& & \hspace{2.0cm}-\,2e^{-\beta\sqrt{q^2-im^2}}\cos(r\cdot\kappa)+1)\,.
\eeq
For $-M^{4}_{r\cdot\kappa}<m^4=-M^4<0$, we find instead
\beq
& & \Delta \hat K_\kappa(m^4,m^4_{\rm vac})=-\,\frac{M^4}{32\pi^2}\left(1+\ln\frac{m^4_{\rm vac}}{M^4}\right)\nonumber\\
& & \hspace{0.5cm}+\,T\int_{q<M}\ln(2\cos(\beta\sqrt{M^2-q^2})-2\cos(r\cdot\kappa))\nonumber\\
& & \hspace{0.5cm}+\,T\int_{q>M}\ln(e^{-2\beta\sqrt{q^2-M^2}}\nonumber\\
& & \hspace{2.5cm}-\,2e^{-\beta\sqrt{q^2-M^2}}\cos(r\cdot\kappa)+1)\nonumber\\
& & \hspace{0.5cm}+\,T\int_q\ln(e^{-2\beta\sqrt{q^2+M^2}}\nonumber\\
& & \hspace{2.0cm}-\,2e^{-\beta\sqrt{q^2+M^2}}\cos(r\cdot\kappa)+1)\,. 
\eeq

\begin{figure}[t]  
\epsfig{file=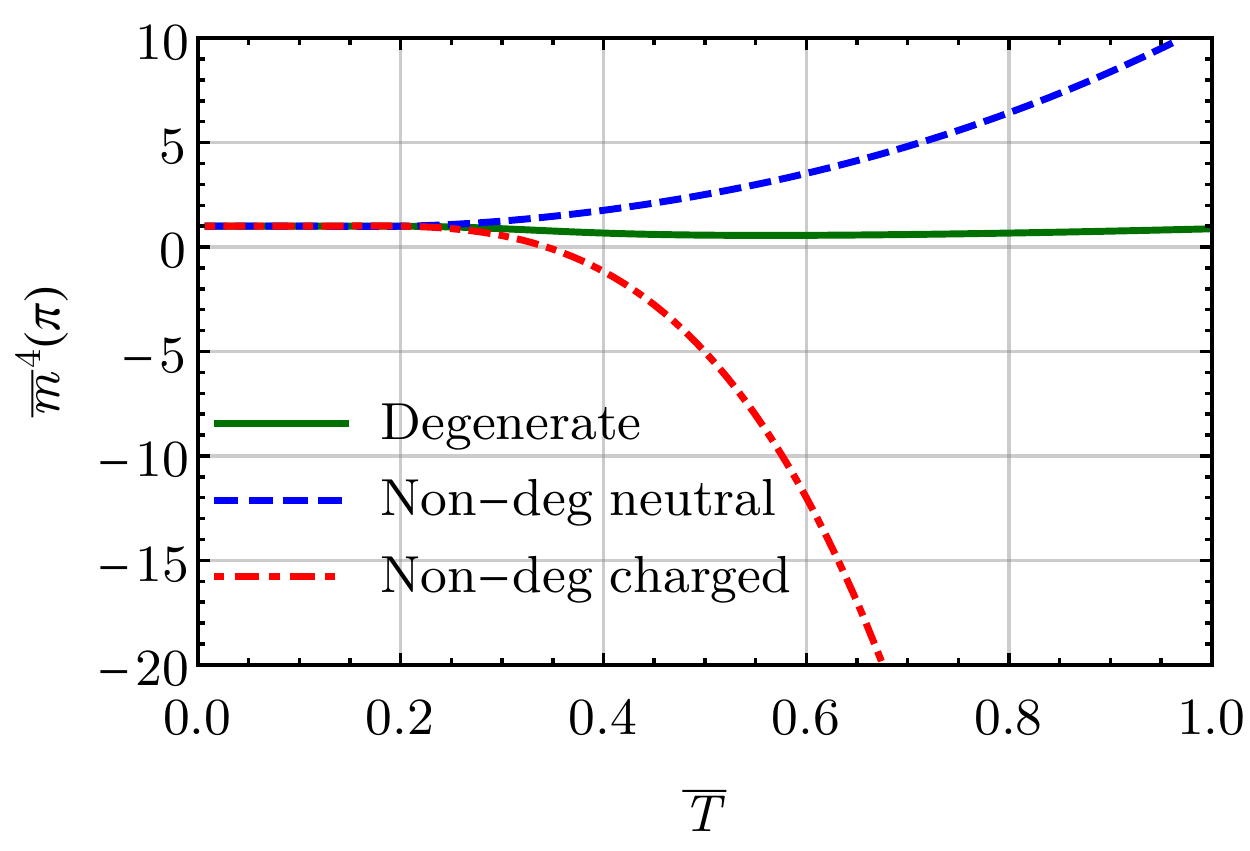,width=7.7cm}\\
\epsfig{file=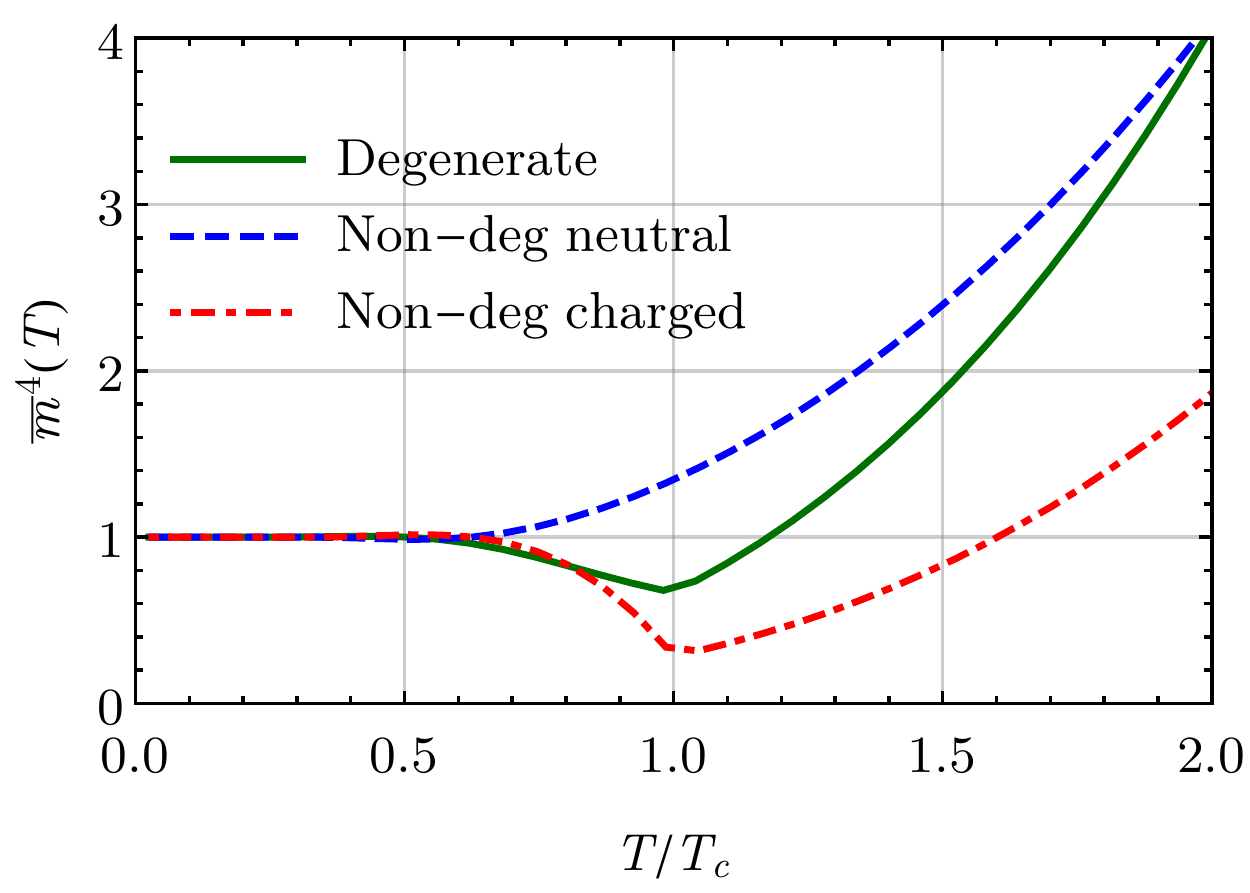,width=7.7cm}
\caption{Top: Degenerate vs non-degenerate Gribov parameters (in units of $m_{\rm vac}$) for a confining background. These correspond to the actual Gribov parameters up to $\bar T_{\rm c}\equiv T_{\rm c}/m_{\rm vac}\sim 0.402$ and $\bar T_{\rm c}\sim 0.324$ respectively. Bottom: Gribov parameters at the minimum of the background effective potential.}\label{fig:Gribov_params_conf_SU(2)}
\end{figure}

%%%%%
\section{Application to the deconfinement transition}\label{sec:numerics}
In what follows we use the previous formalism to study the deconfinement transition in SU(2) and SU(3) Yang-Mills theories. We minimize the background effective potential with respect to the order parameter $r$, taking into account the $r$-dependence of the Gribov parameter(s) via the gap equation(s), that is by minimizing $V(r,\{m^4_\kappa(r)\})$. We first revisit the SU(2) results of Ref.~\cite{Canfora:2015yia} by including the possibility of color dependent Gribov parameters and then extend our analysis to the SU(3) case.

%%%
\subsection{SU(2) case}
In this case $\kappa\in\{-1,0,+1\}$ and the confining point corresponds to $r=\pi$. The partially degenerate and non-degenerate cases coincide.

\subsubsection{Critical temperature}
 Since we expect the transition to be second order, we can evaluate $T_{\rm c}$ by requiring that (we illustrate the degenerate case here but the same discussion holds for the non-degenerate one)
\beq
\left.\frac{d^2}{dr^2}V(r,m^2(r))\right|_{r=\pi}=0\,.
\eeq
Since $\partial V/\partial m^2|_{r,m^2(r)}=0$, we have
\beq\label{eq:45}
\frac{d}{dr}V(r,m^2(r))=\frac{\partial}{\partial r}V(r,m^2(r))
\eeq
and then
\beq\label{eq:46}
\frac{d^2}{dr^2}V(r,m^2(r)) & = & \frac{\partial^2}{\partial r^2}V(r,m^2(r))\nonumber\\
& + & \frac{\partial^2}{\partial r\partial m^2}V(r,m^2(r))\frac{dm^2(r)}{dr}\,.
\eeq
Finally, it is easily shown that $dm^2(r)/dr|_{r=\pi}=0$.\footnote{This is because $dm^2(r)/dr|_{r=\pi}$ is proportional to
\beq\sum_\alpha \alpha\int_Q^T \frac{Q^\alpha_0Q_\alpha^2}{Q^4_\alpha+m^4} & = & \int_Q^T \frac{(\omega_n+\pi T)((\omega_n+\pi T)^2+q^2)}{((\omega_n+\pi T)^2+q^2)^2+m^4}\nonumber\\
& - & \int_Q^T \frac{(\omega_n-\pi T)((\omega_n-\pi T)^2+q^2)}{((\omega_n-\pi T)^2+q^2)^2+m^4}.\nonumber
\eeq
Using the changes of variables $\omega_n\to-\omega_n-2\pi T$ and $\omega_n\to-\omega_n+2\pi T$, we find that the sum-integrals are both zero.} Therefore
\beq
\left.\frac{d^2}{dr^2}V(r,m^2(r))\right|_{r=\pi} & = & \left.\frac{\partial^2}{\partial r^2}V(r,m^2(r))\right|_{r=\pi}\,.
\eeq
After a simple calculation, the condition for a vanishing curvature reads
\beq
\frac{3}{2}\,{\rm Re}\int_q\!\frac{e^{-\beta\sqrt{q^2+im^2(\pi)}}}{(e^{-\beta\sqrt{q^2+im^2(\pi)}}+1)^2}=\int_q\!\frac{e^{-\beta q}}{(e^{-\beta q}+1)^2}\,.
\eeq
The non-degenerate case is obtained upon making the replacement $m^2(\pi)\to m^2_{\rm Ch}(\pi)$.

\begin{figure}[t]  
\epsfig{file=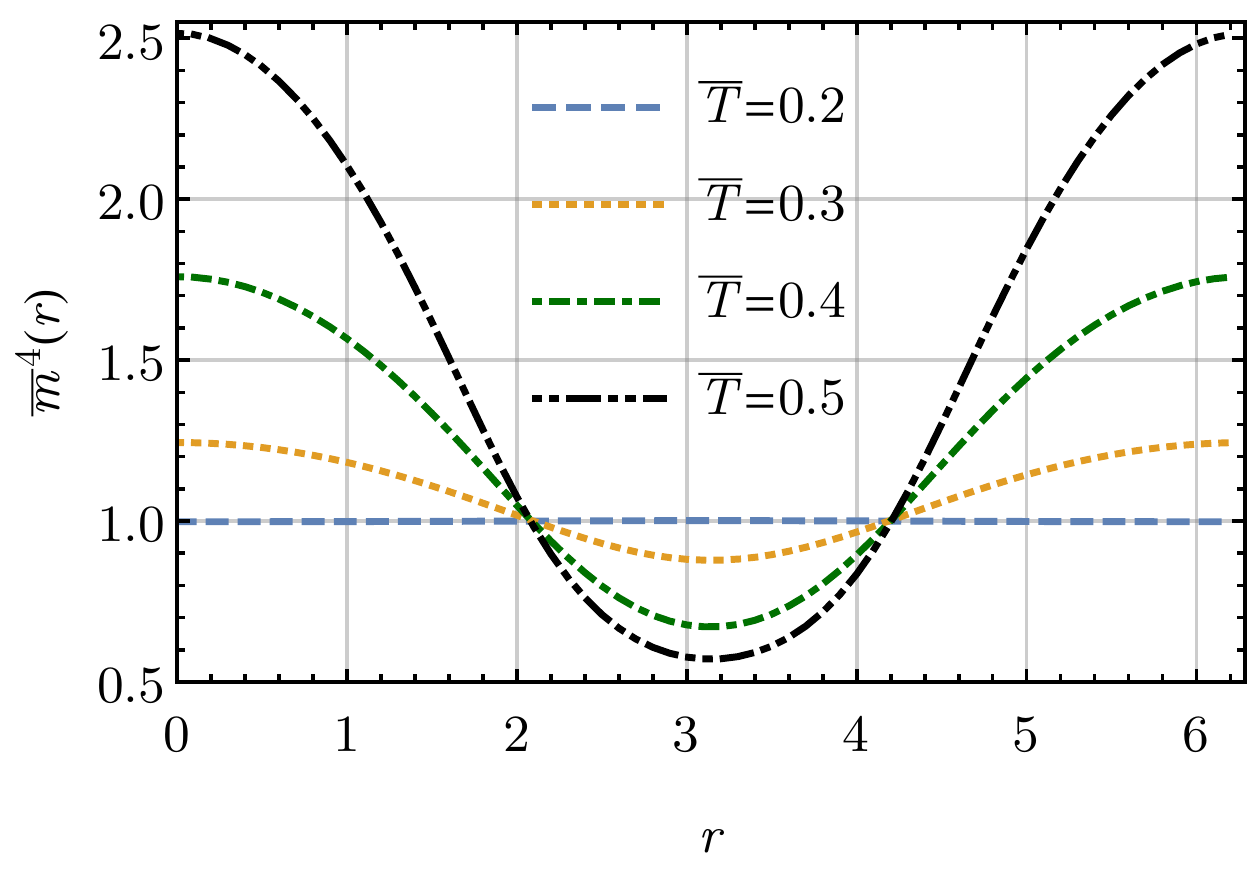,width=8cm}\\
\epsfig{file=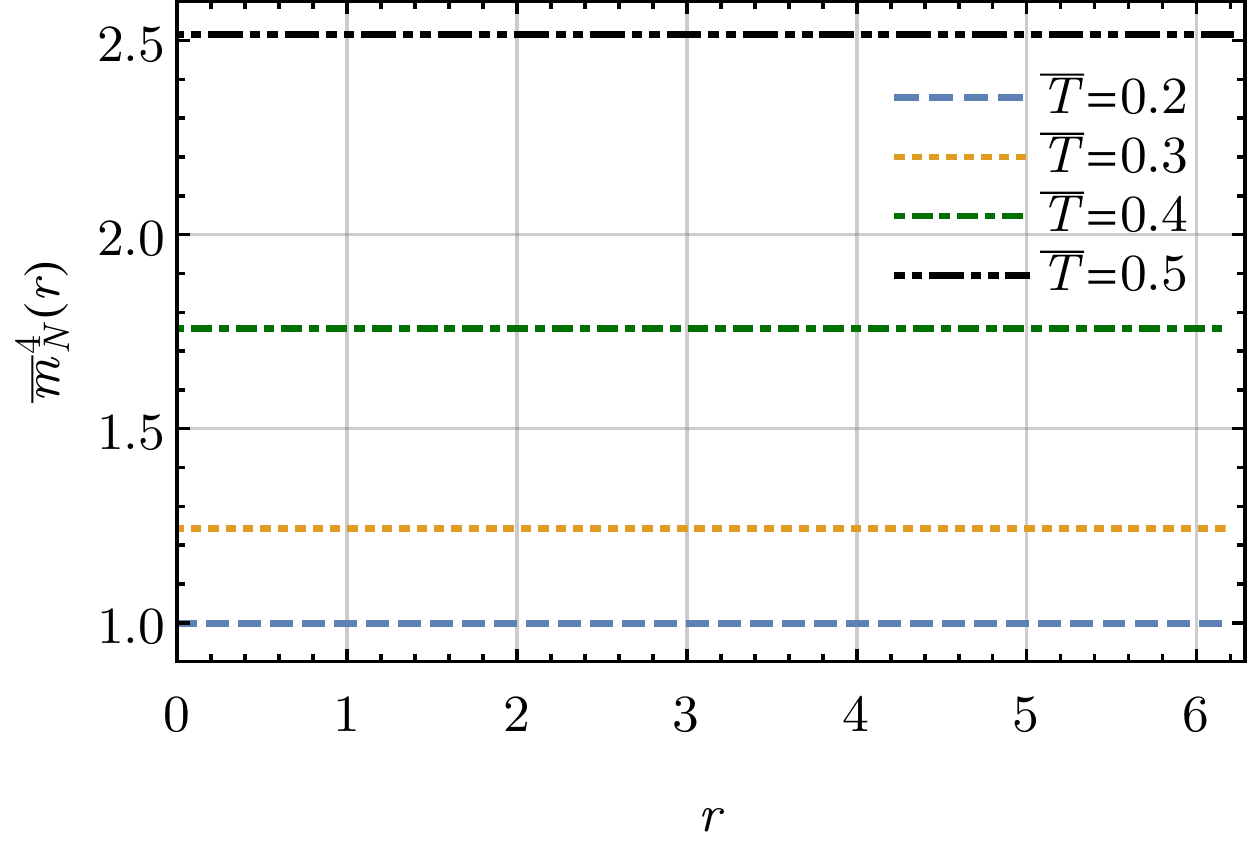,width=8cm}\\
\epsfig{file=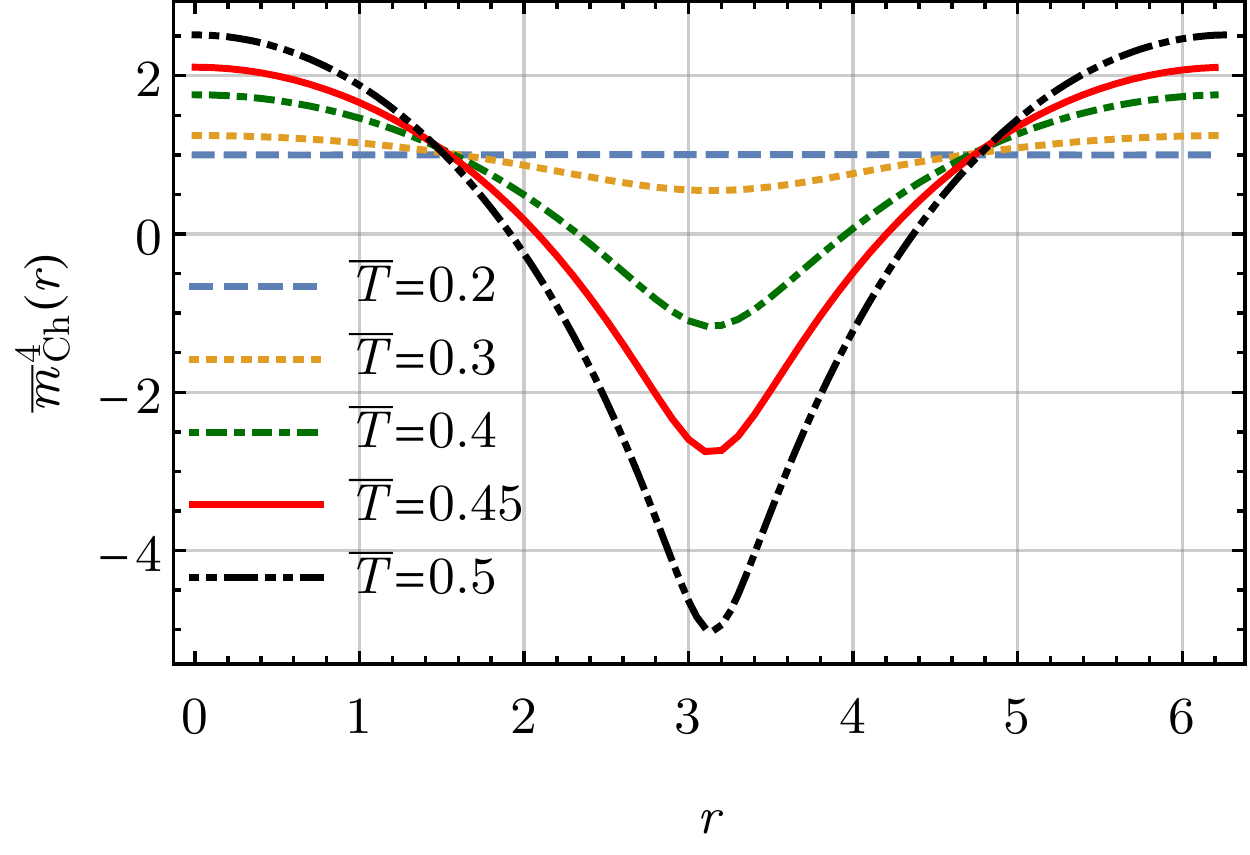,width=8cm}
\caption{$r$-dependence of the Gribov parameters for various temperatures (in units of $m_{\rm vac}$). Top: degenerate case. Middle: non-degenerate case, neutral mode. Bottom: non-degenerate case, charged mode.}\label{fig:mofr}
\end{figure}

\begin{figure}[t]  
\epsfig{file=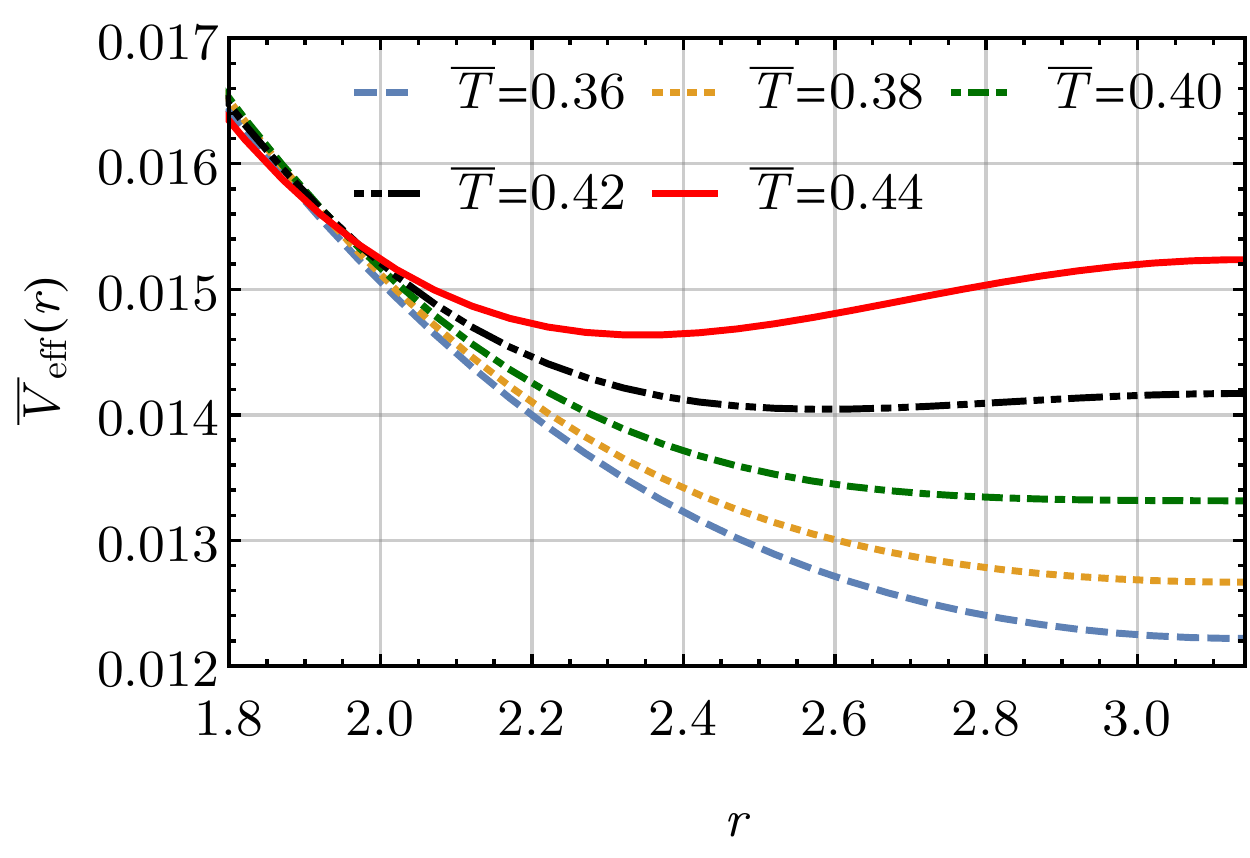,width=8cm}
\,\,\,\,\,\epsfig{file=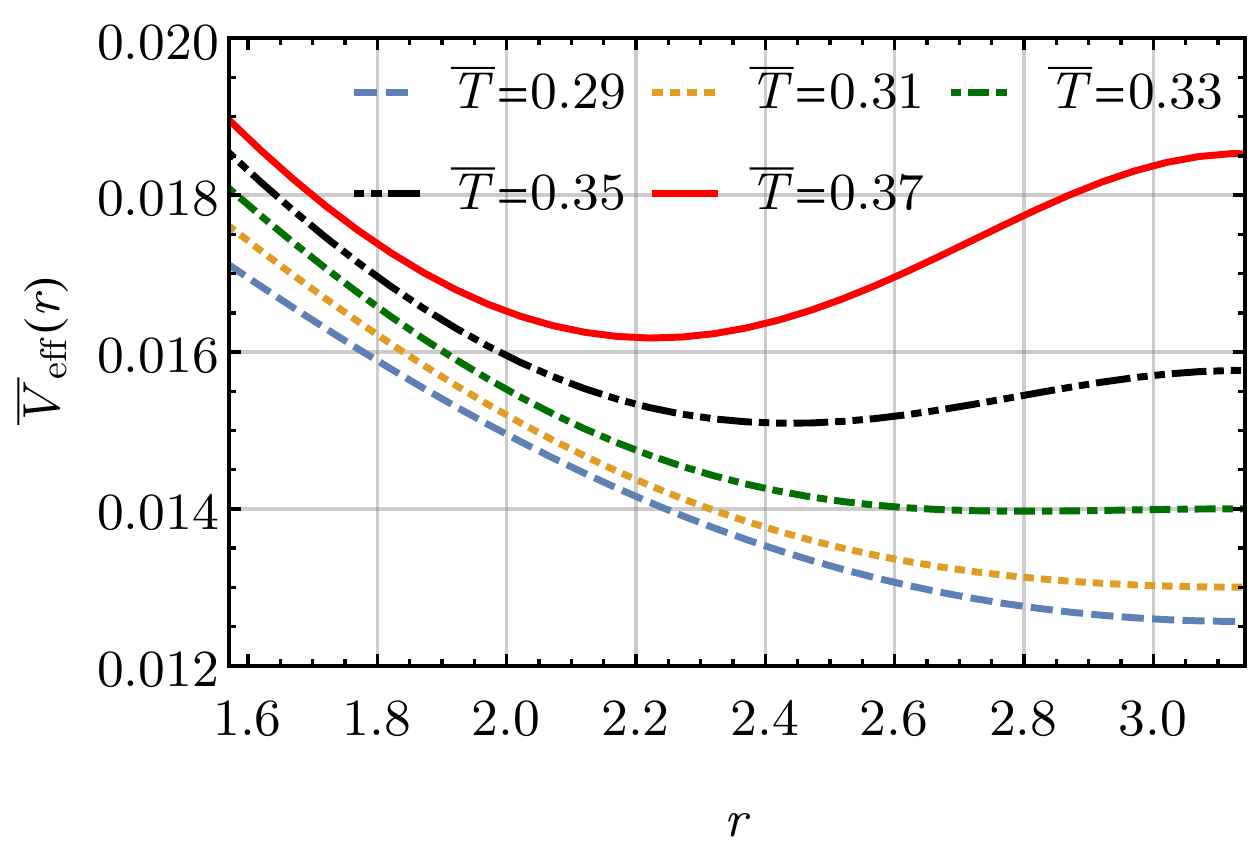,width=8cm}
\caption{SU(2) background effective potentials for various temperatures (in units of $m_{\rm vac}$). Top: degenerate case. Bottom: non-degenerate case.}\label{fig:pots_SU2}
\end{figure}

In order to find the transition temperatures in each case, we need to determine the temperature dependence of $m^4(\pi)$ and $m^4_{\rm Ch}(\pi)$. This is shown in Fig.~\ref{fig:Gribov_params_conf_SU(2)}, together with the temperature dependence of $m^4_{\rm N}(\pi)$ for completeness. We observe that $m^4_{\rm Ch}(\pi)$ decreases rapidly and even changes sign (as already anticipated in the previous section) at a temperature $T/m_{\rm vac}\sim 0.344$,  obtained from solving Eq.~(\ref{eq:T0}) which takes here the form
\beq
\frac{1}{8\pi^2}=\int_q \left[\frac{1-2 f_{q}}{4q^3}+\frac{1}{m^2_{\rm vac}}{\rm Im}\frac{1}{2\sqrt{q^2+im^2_{\rm vac}}}\right].\nonumber
\eeq
The decrease of $m^4_{\rm Ch}(\pi)$ with the temperature has the effect of lowering the transition temperature as compared to the degenerate case. We find
\beq
\frac{T_{\rm c}^{\rm non-deg}}{m_{\rm vac}}\sim 0.324\,,
\eeq
which should be compared to the result of Ref.~\cite{Canfora:2015yia}
\beq
\frac{T_{\rm c}^{\rm deg}}{m_{\rm vac}}\sim 0.402\,.
\eeq
This represents a change of the transition temperature by 20\%--25\%.

\subsubsection{Effective potential}
In order to compute the potential as a function of $r$, we first need to determine,  for each temperature, the $r$-dependence of the Gribov parameters. This dependence is shown in Fig.~\ref{fig:mofr}. Above $T/m_{\rm vac}\sim 0.344$, a gap opens in the values of $r$, over which $m^4_{\rm Ch}$ becomes negative. At each temperature, the boundaries of this interval can be determined by solving
\beq
\frac{1}{8\pi^2}=\int_q \left[\frac{1+2 {\rm Re}\,n_{q-ir\cdot\kappa T}}{4q^3}+\frac{1}{m^2_{\rm vac}}{\rm Im}\frac{1}{2\sqrt{q^2+im^2_{\rm vac}}}\right].\nonumber
\eeq
We stress that, despite $m^4_{\rm Ch}$ becoming negative, the potential remains real. The results for the potential are shown in Fig.~\ref{fig:pots_SU2}. We verify that the transition is second order and that the transition temperatures agree with the estimates given above. We also note that the minimum never enters the region of negative $m^4_{\rm Ch}$, as can also be seen in Fig.~\ref{fig:Gribov_params_conf_SU(2)} (bottom), where we show the Gribov parameters at the minimum of the potential.

\begin{figure}[t]  
\epsfig{file=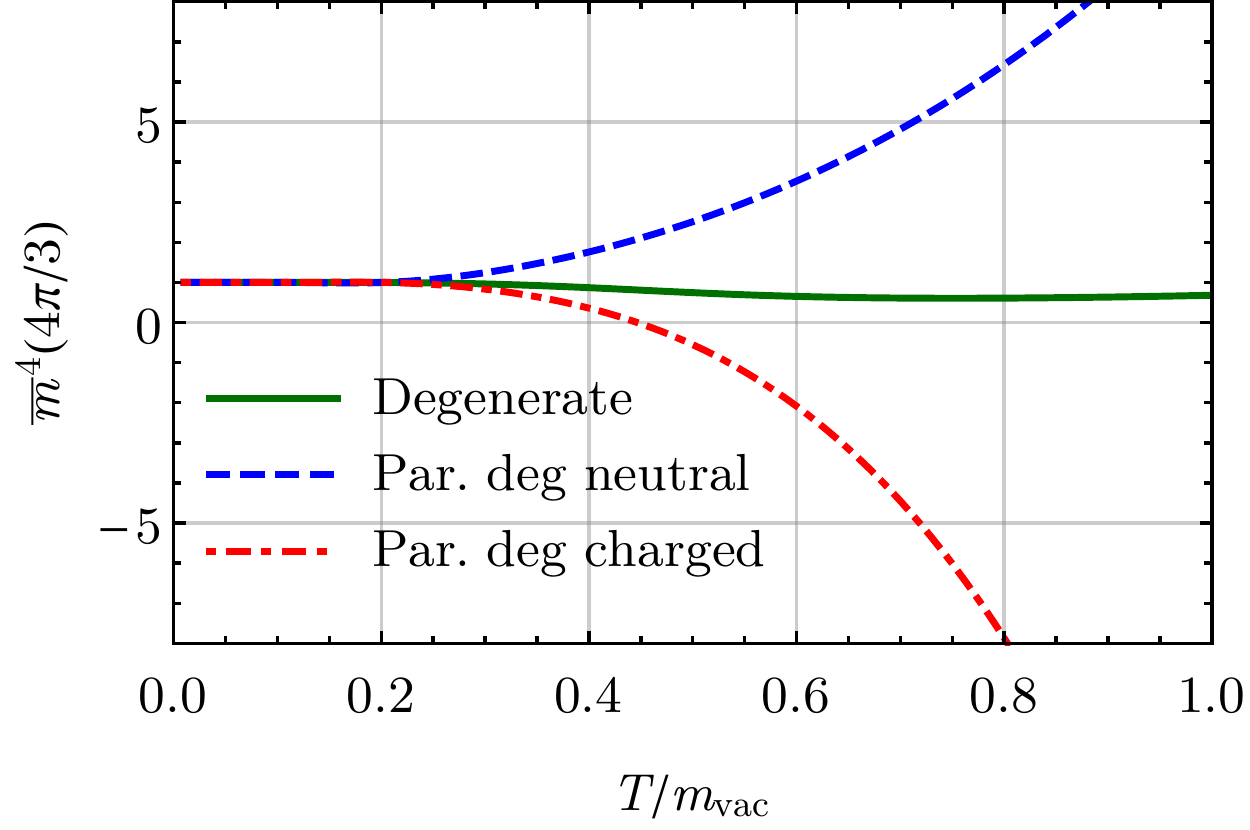,width=7.7cm}\\
\epsfig{file=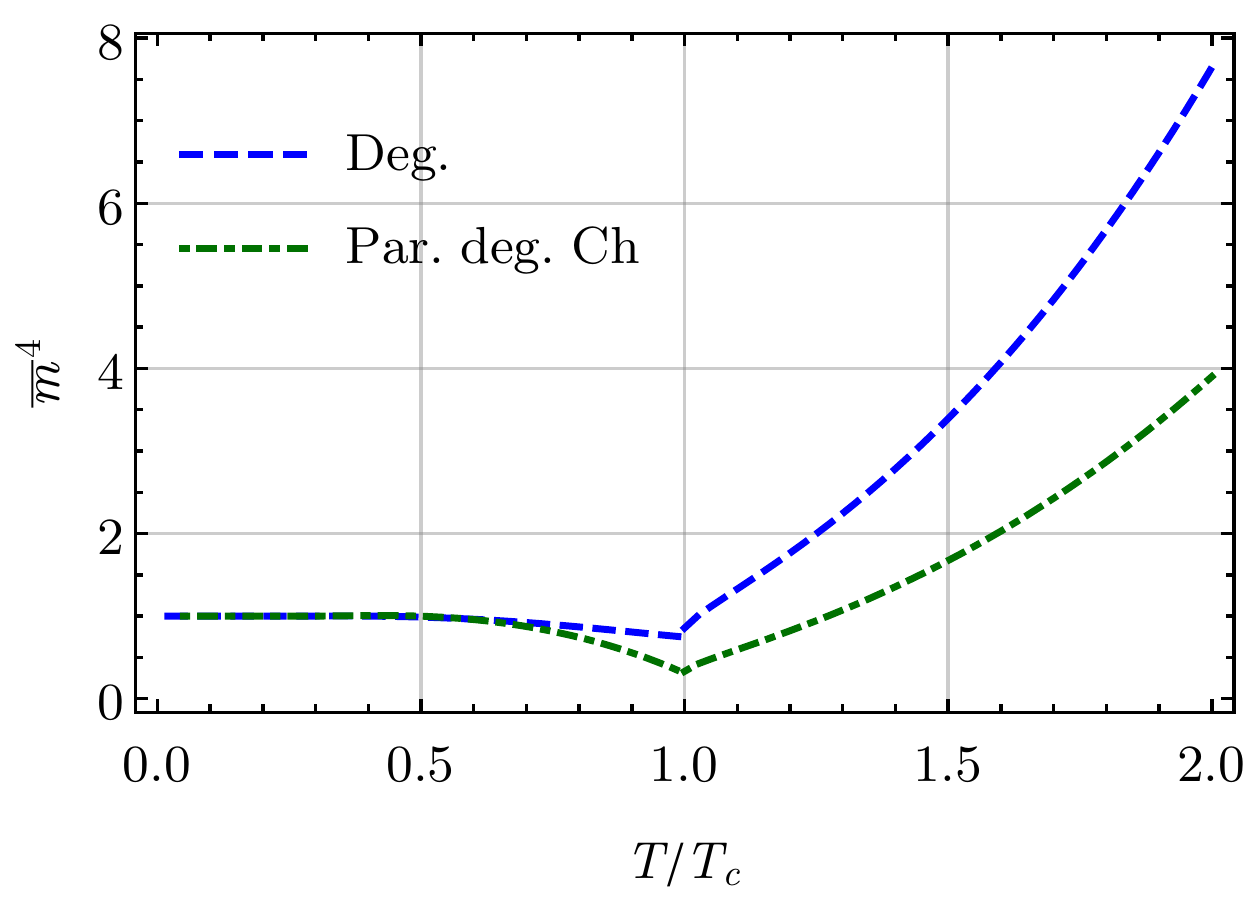,width=7.7cm}\\
\epsfig{file=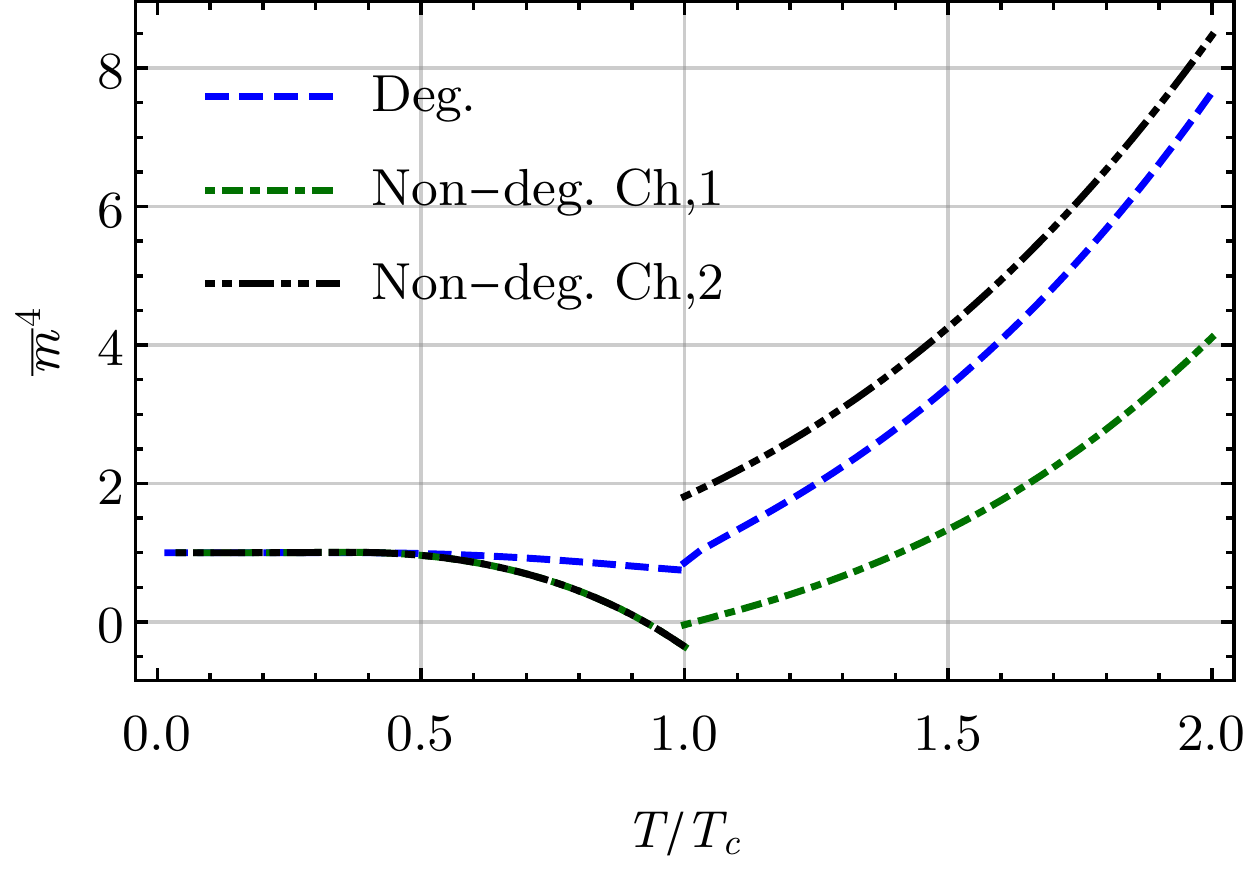,width=7.7cm}
\caption{Top: Degenerate vs partially or non-degenerate Gribov parameters (in units of $m_{\rm vac}$) for a confining background. Middle: Degenerate vs partially degenerate Gribov parameters at the minimum of the background effective potential. Bottom: Degenerate vs non-degenerate Gribov parameters at the minimum of the background effective potential.}\label{fig:Gribov_params_conf_SU(3)}
\end{figure}

%%%
\subsection{SU(3) case}
We can repeat a similar analysis for the SU(3) gauge group. In this case there are two neutral modes $\smash{\kappa=0^{(3)}}$ and $\smash{\kappa=0^{(8)}}$, and six roots $\smash{\kappa=\alpha}$, with $\alpha\in\{\pm(1,0),\pm(1/2,\sqrt{3}/2),\pm(1/2,-\sqrt{3}/2)\}$. The confining point is $r=(4\pi/3,0)$. Moreover, due to charge conjugation invariance, we can restrict the analysis to $r=(r_3,0)$. We shall rename $r_3$ as $r$ in what follows. We also mention that 
\beq
m^4_{(1,0)}=m^4_{(-1,0)}
\eeq 
and
\beq
m^4_{(1/2,\sqrt{3}/2)} & = & m^4_{(-1/2,\sqrt{3}/2)}\nonumber\\
& = & m^4_{(1/2,-\sqrt{3}/2)}=m^4_{(-1/2,-\sqrt{3}/2)}\,.
\eeq 
Therefore, in the non-degenerate case, we only need to introduce two charged Gribov parameters, denoted $m^4_{\rm Ch,1}$ and $m^4_{\rm Ch,2}$ respectively. As it is easily checked, at the confining point they both coincide with the charged Gribov parameter $m^4_{\rm Ch}(r)$ of the partially degenerate case, and in general, $m^4_{\rm Ch,2}(r)=m^4_{\rm Ch,1}(r/2)$.

\begin{figure}[t]  
\epsfig{file=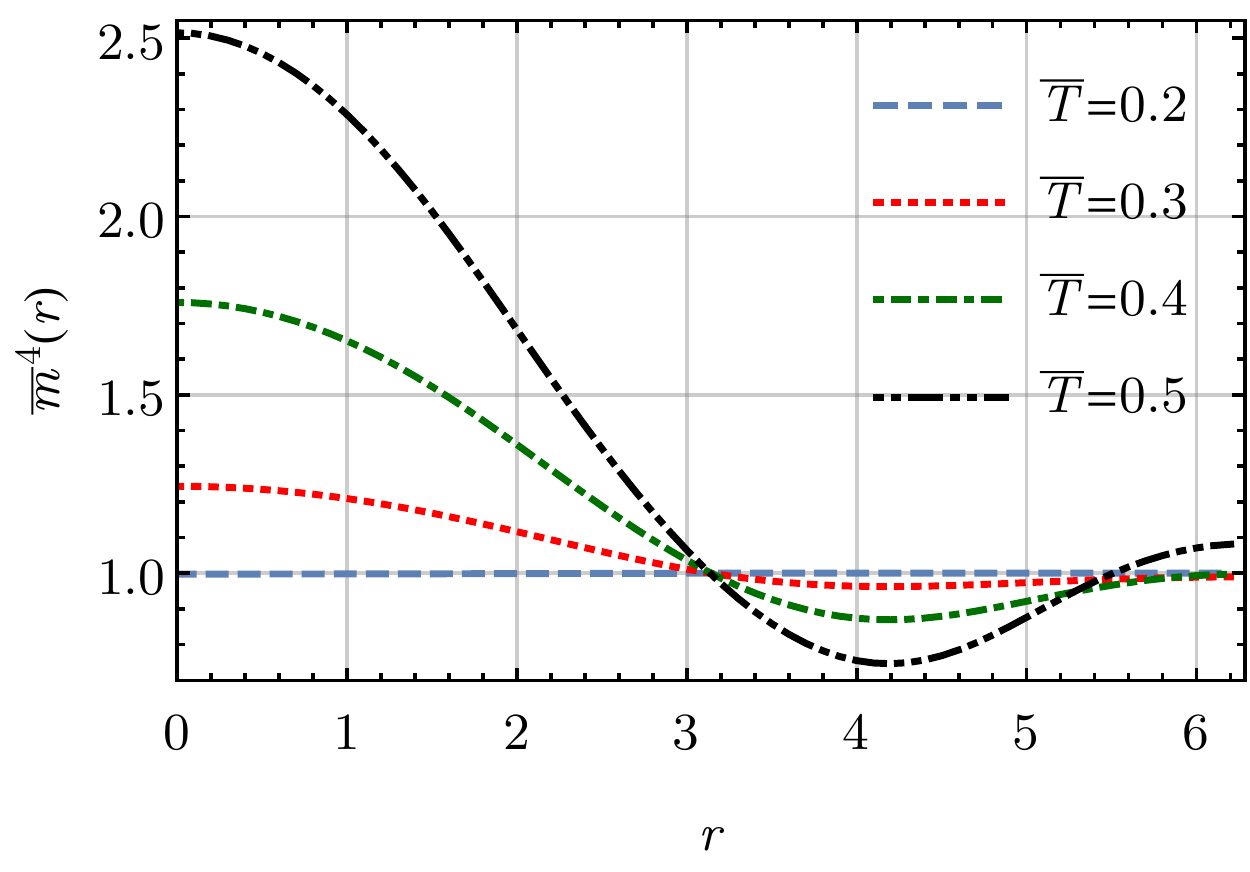,width=8cm}\\
\epsfig{file=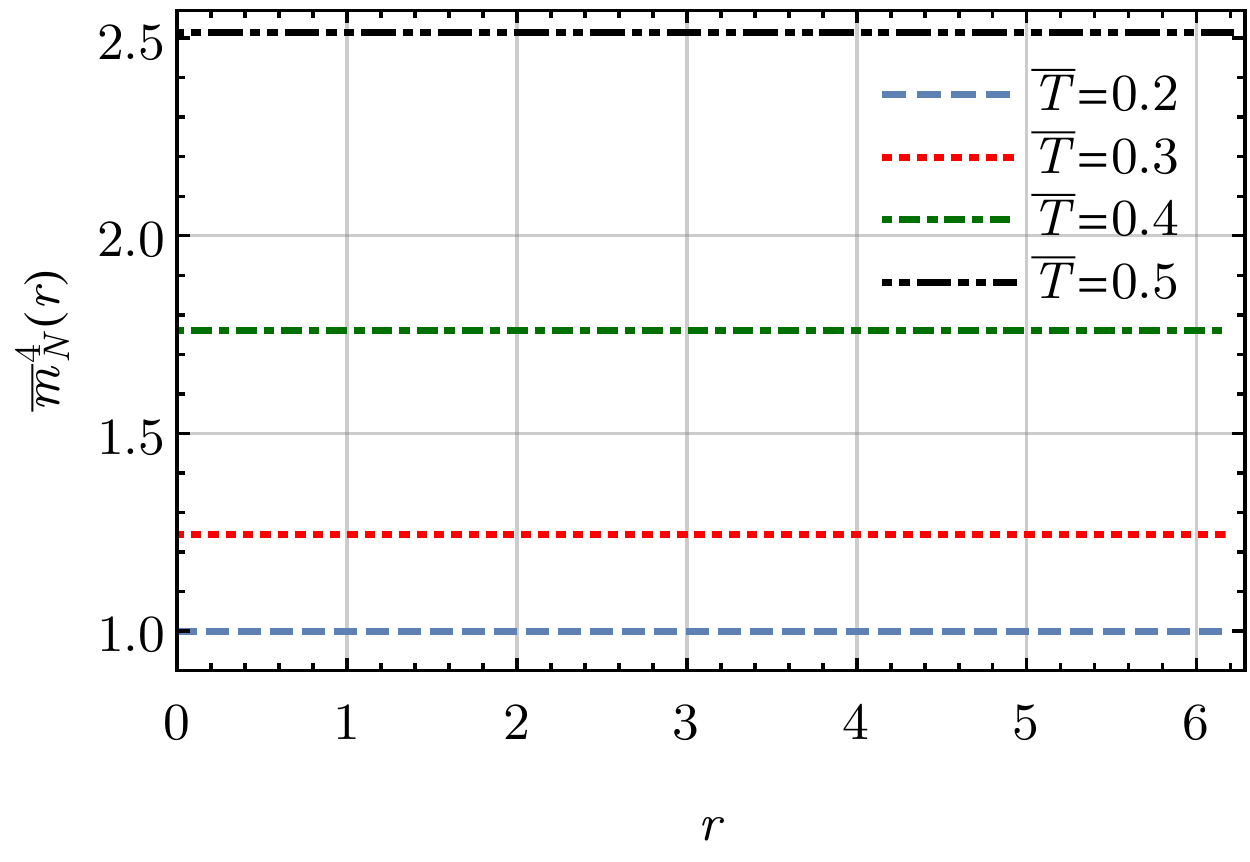,width=8cm}\\
\epsfig{file=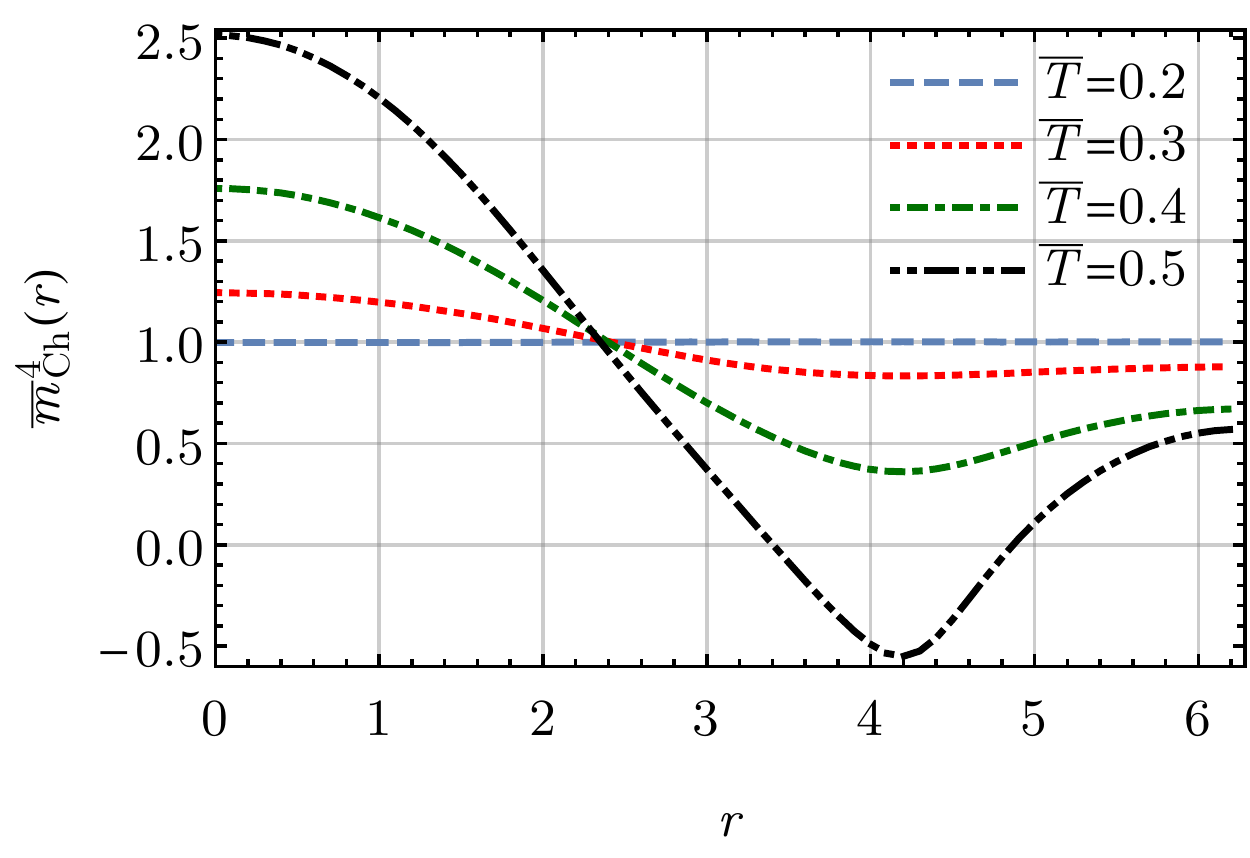,width=8cm}
\caption{$r$-dependence of the Gribov parameters for various temperatures (in units of $m_{\rm vac}$). We show only the degenerate and partially degenerate cases. The non-degenerate charged Gribov parameters are obtained in terms of the SU(2) one respectively as $m^4_{\rm Ch,1}(r)=m^4_{\rm Ch,SU(2)}(r)$ and $m^4_{\rm Ch,2}(r)=m^4_{\rm Ch,SU(2)}(r/2)$.}\label{fig:mofr_SU3}
\end{figure}

\begin{figure}[t]  
\epsfig{file=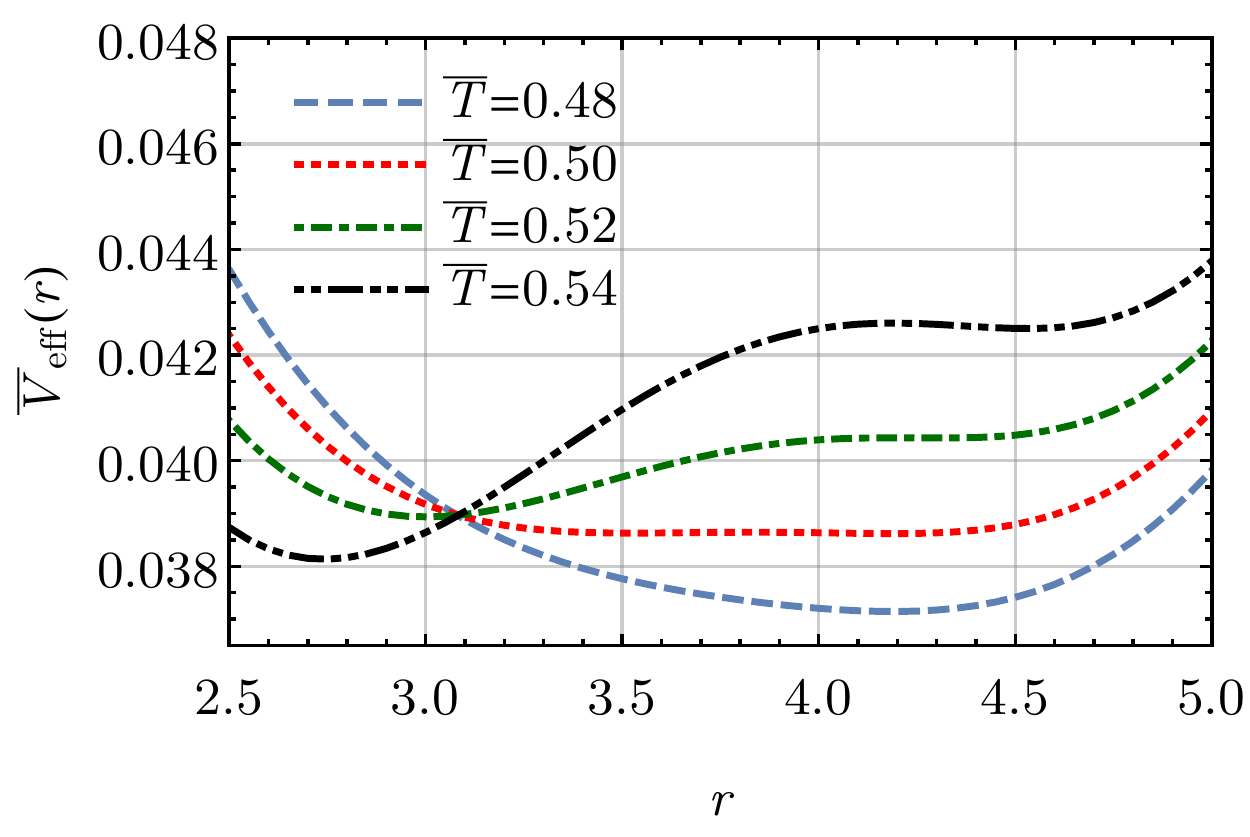,width=8cm}
\,\,\,\,\,\epsfig{file=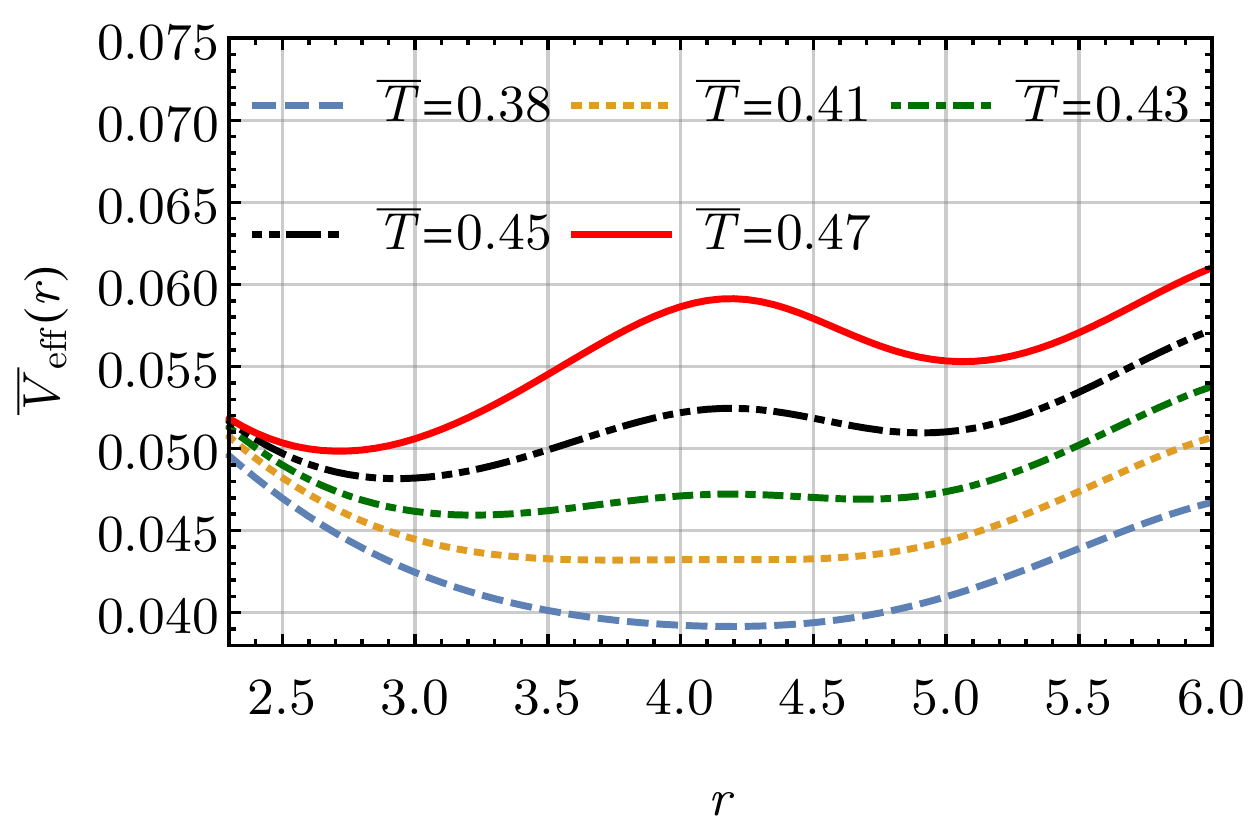,width=8cm}
\,\,\,\,\,\epsfig{file=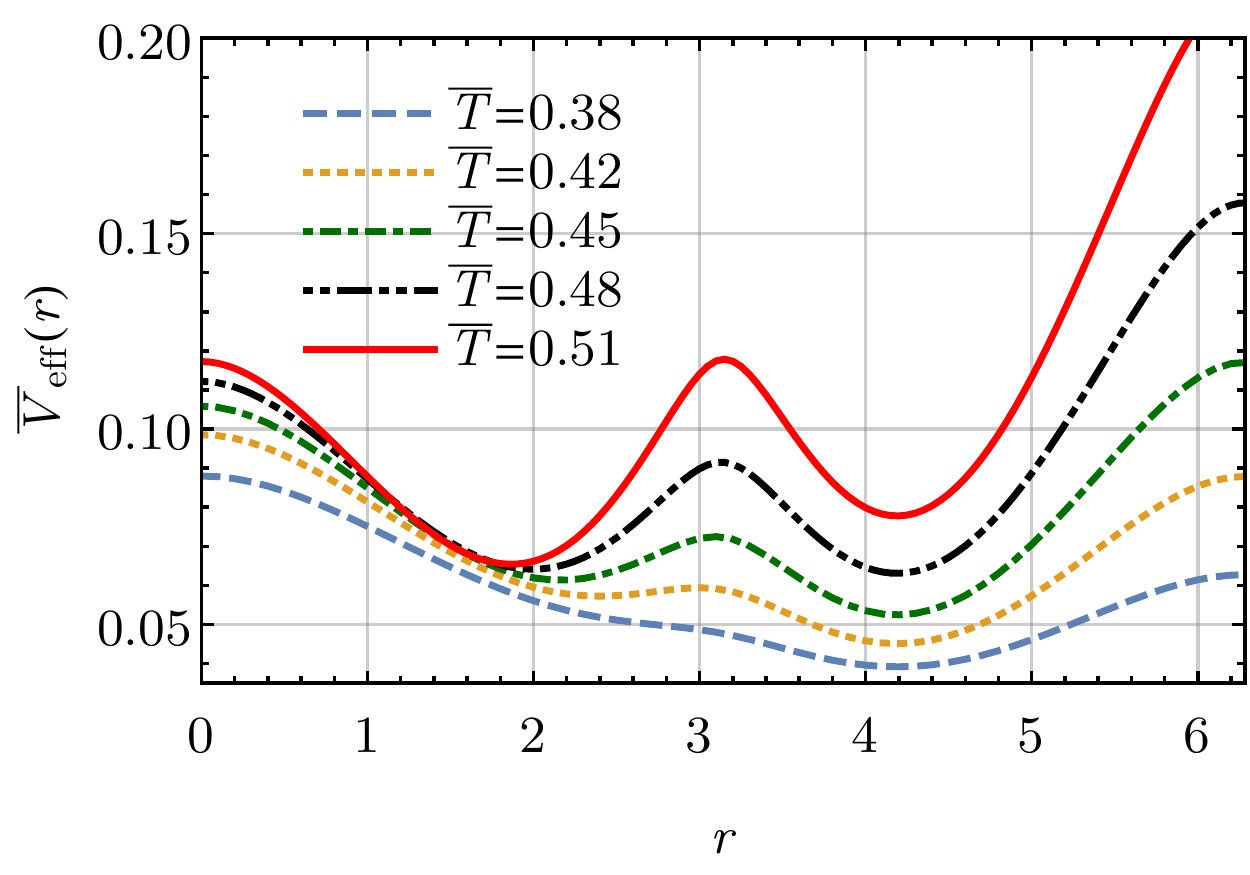,width=8cm}
\caption{SU(3) background effective potentials for various temperatures (in units of $m_{\rm vac}$) with degenerate, partially degenerate and non-degenerate Gribov parameters.}\label{fig:pots_SU3}
\end{figure}

\subsubsection{Highest spinodal}
In the SU(3) case, we expect the transition to be first order so we cannot determine the transition temperature so simply as above. However, we expect the spinodal temperatures to be quite close to the transition temperature. The highest spinodal can be determined using the same method as above because it occurs at $r=4\pi/3$. We first evaluate the curvature at $r=4\pi/3$. To this purpose, we notice that Eqs.~(\ref{eq:45}) and (\ref{eq:46}) are still valid. Moreover, both in the degenerate and the partially degenerated cases, it is easily shown that $dm^2(r)/dr|_{r=4\pi/3}=0$.\footnote{This is because $dm^2(r)/dr|_{r=4\pi/3}$ is proportional to
\beq
\sum_\alpha \alpha_3\int_Q^T \frac{Q^\alpha_0Q_\alpha^2}{Q^4_\alpha+m^4_{\rm Ch}} & \!=\! & 2\left[\int_Q^T \frac{(\omega_n+4\pi/3T)(\omega_n+4\pi/3T)^2+q^2)}{((\omega_n+4\pi/3T)^2+q^2)^2+m^4_{\rm Ch}}\right.\nonumber\\
& \!+\! & \left.\int_Q^T \frac{(\omega_n+2\pi/3T)(\omega_n+2\pi/3T)^2+q^2)}{((\omega_n+2\pi/3T)^2+q^2)^2+m^4_{\rm Ch}}\right].\nonumber
\eeq
Using the change of variables $\omega_n\to-\omega_n-2\pi T$ in the second integral, we find that the bracket is zero.} It follows that
\beq
\left.\frac{d^2}{dr^2}V(r,m(r))\right|_{r=4\pi/3}\!\!\!\!=\left.\frac{\partial^2}{\partial r^2}V(r,m(r))\right|_{r=4\pi/3}\,.
\eeq
In the degenerate case, the condition for a vanishing curvature reads then
\beq
& & \frac{3}{2}\,{\rm Re}\int_q\frac{e^{-3\beta\sqrt{q^2+im^2}}+4e^{-2\beta\sqrt{q^2+im^2}}+e^{-\beta\sqrt{q^2+im^2}}}{(e^{-2\beta\sqrt{q^2+im^2}}+e^{-\beta\sqrt{q^2+im^2}}+1)^2}\nonumber\\
& & \hspace{1.0cm} -\,\int_q \frac{e^{-3\beta q}+4e^{-2\beta q}+e^{-\beta q}}{(e^{-2\beta q}+e^{-\beta q}+1)^2}=0\,.
\eeq
The partially degenerate case is obtained upon making the replacement $m^2(4\pi/3)\to m^2_{\rm Ch}(4\pi/3)$. The non-degenerate case cannot be treated in this way. The corresponding transition temperature will be determined in the next section.

The temperature dependence of the Gribov parameters at the minimum is shown in Fig.~\ref{fig:Gribov_params_conf_SU(3)}. Using this temperature dependence, we can determine the spinodal temperatures. We find
\beq
\frac{T_{\rm c}^{\rm part-deg}}{m_{\rm vac}}\simeq\frac{T_{\rm spinod}^{\rm part-deg}}{m_{\rm vac}}\sim 0.409\,,
\eeq
as compared to the result of Ref.~\cite{Canfora:2015yia}
\beq
\frac{T_{\rm c}^{\rm deg}}{m_{\rm vac}}\simeq\frac{T_{\rm spinod}^{\rm deg}}{m_{\rm vac}}\sim 0.512\,,
\eeq
so again a $20\%$ difference.

\subsubsection{Effective potential}
Once again, to compute the potential we need to know the background dependence of the Gribov parameters. This is shown in Fig.~\ref{fig:mofr_SU3} where one sees that the charged ones can becomes negative. For the degenerate and partially degenerate cases, we find transition temperatures very close to the higher spinodal temperatures determined above. For the completely non-degenerate case, we find
\beq
\frac{T_{\rm c}^{\rm non-deg}}{m_{\rm vac}}\sim 0.48\,,
\eeq
which represents a $6\%$ difference with respect to the degenerate case. We mention that, as compared to the degenerate and partially degenerated cases, it was crucial in the non-degenerate case to be able to resolve the potential in the region where the Gribov parameters become negative because the minimum lies in this region just before the transition occurs, as can be seen in the bottom plot of Fig.~\ref{fig:Gribov_params_conf_SU(3)}.

%%%
\subsection{Comparison with the Curci-Ferrari model}
We finally compare our model at one-loop with a similar calculation in the CF model. To this purpose we show the Polyakov loops in Fig.~\ref{fig:pol_SU3}. We observe that the growth of the order parameter above $T_{\rm c}$ is slower in the Gribov-Zwanziger approach than in the CF model. This is more qualitatively in line with the behavior observed on the lattice.

We shall not display the thermodynamical observables in the low temperature phase since they suffer from problems similar to those reported in other approaches \cite{Reinosa:2014zta,Canfora:2015yia,Quandt:2017poi}, specially in the limit of vanishing temperature. At the transition however, we can estimate the latent heat which, at one-loop order, does not depend on the parameter $m_{\rm vac}$. We find $(L/T_c^4)\approx 0.31$ in the degenerate case, $(L/T_c^4)\approx 0.17$ in the partially degenerate case, and $(L/T_c^4)\approx 3.25$ in the non-degenerate case, to be compared to the value obtained $(L/T_c^4)\approx 0.43$ obtained within the Curci-Ferrari model at one-loop \cite{Reinosa:2015gxn}. The lattice gives instead $(L/T_c^4)\approx 1.4$ \cite{Beinlich:1996xg}. It would be interesting to see if higher order corrections can help diminishing the discrepancy in at least one of the scenarios.

\begin{figure}[h]  
\epsfig{file=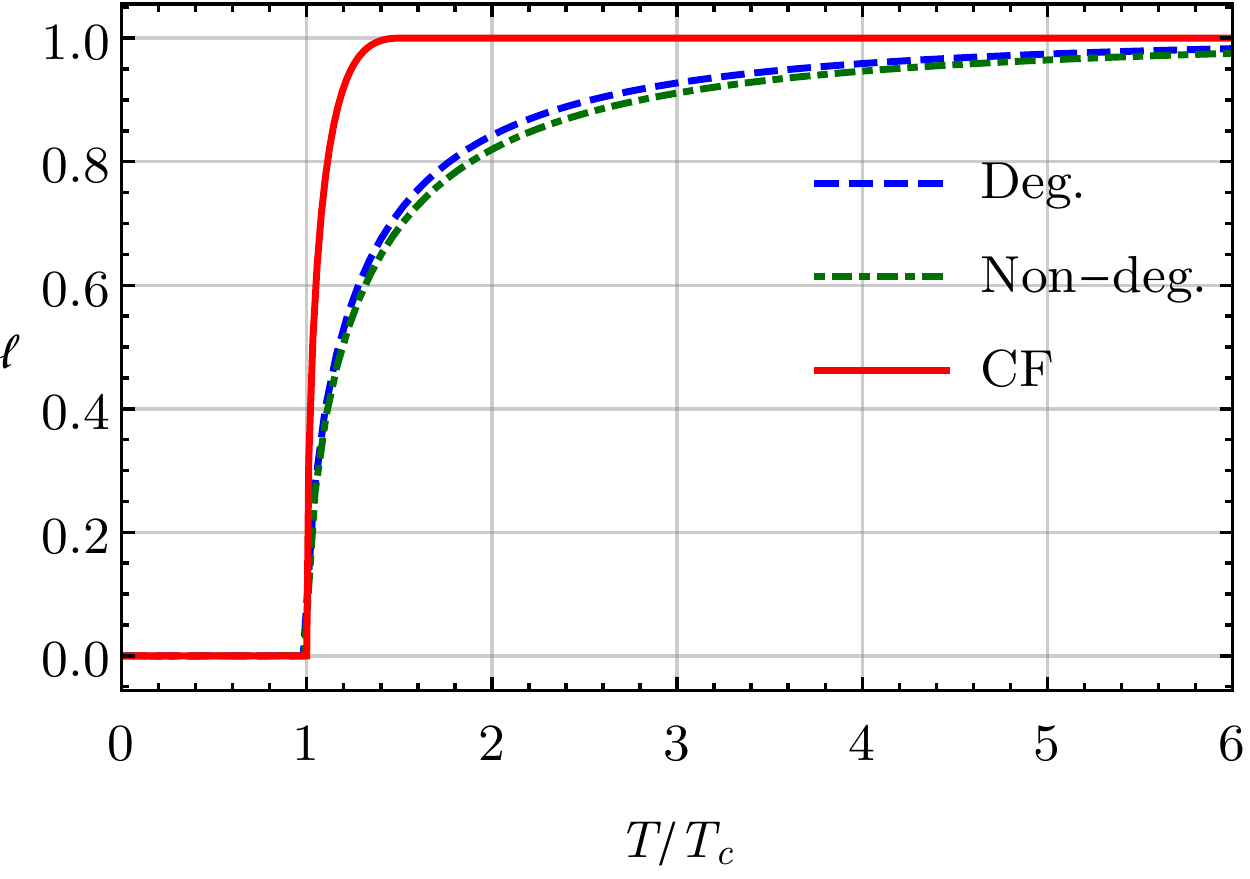,width=8cm}
\,\,\,\epsfig{file=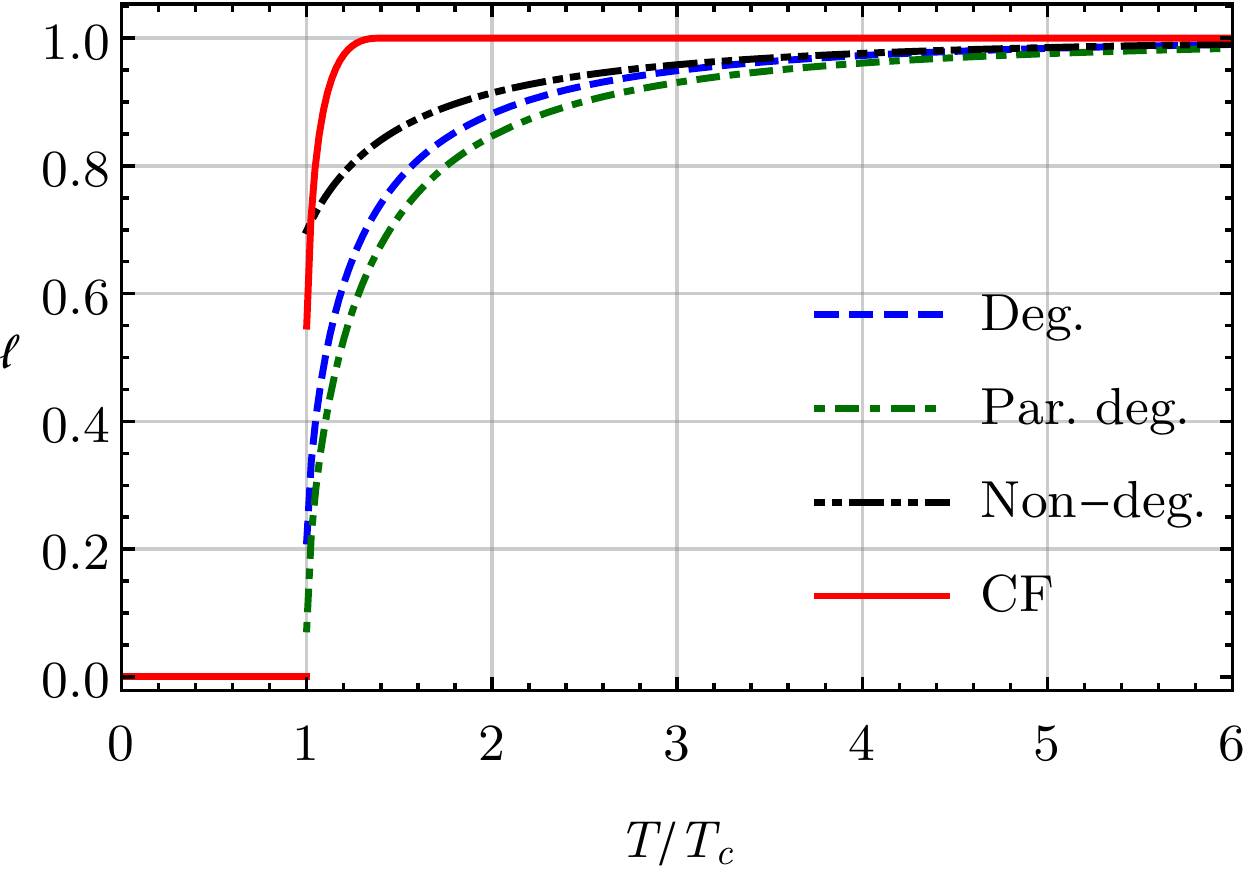,width=8cm}
\caption{Polyakov loops. Top: SU(2). Bottom: SU(3).}\label{fig:pol_SU3}
\end{figure} 

%%%%%
\section{Relation with the Gribov restriction}\label{sec:Gribov}
In this section, we investigate the relation between the model (\ref{eq:action_prop_upgrade_2}) and the restriction of the functional integral to the first Gribov region. We first show that, at zero temperature and to one-loop accuracy, the model can be related Gribov no-pole condition applied to the Landau-DeWitt gauge. We then argue that the result is not so surprising since, at zero temperature, there is a trivial mapping between the Landau and Landau-DeWitt gauges. Finally, we investigate the extension to the finite temperature case, emphasizing similar difficulties than the ones discussed in Refs.~\cite{Cooper:2015sza,Cooper:2015bia}.

%%%
\subsection{Relation with the Gribov no-pole condition\\ at zero temperature}
We first recall how the no-pole condition is constructed at one-loop order in the Landau gauge at zero temperature\footnote{Up to some slight modifications, we follow the nice presentation given in Ref.~\cite{Vandersickel:2011zc}} and then extend it to the Landau-DeWitt gauge.

Consider the ghost propagator ${\cal G}_{ab}(K,P,A)$ in the presence of a gauge field configuration $A$. If we evaluate this propagator for $P=K$ and $b=a$, we obtain
\beq
{\cal G}_{aa}(K,K,A)=\!\!\int_x\int_y (e^{iKx}\delta_{ac})^*(-\partial D)^{-1}_{cd}(x,y)(e^{iKy}\delta_{ad})\,.\nonumber\\
\eeq
If $A$ belongs to the first Gribov region, it follows by construction that ${\cal G}_{aa}(K,K,A)>0$, $\forall a$ and $\forall K$. In other words, by imposing these inequalities, one restricts $A$ to lie in a domain that still contains the Gribov region. Moreover, if starting from inside the Gribov region (say from $A=0$), we approach its boundary (the so-called Gribov horizon), at least one of the ${\cal G}_{aa}(K,K,A)$'s diverges and changes sign. This means that the Gribov horizon lies inside the boundary of the region defined by the conditions ${\cal G}_{aa}(K,K,A)>0$, $\forall a$ and $\forall K$.

In practice, it is not simple to impose the conditions for all $a$'s and $K$'s separately and instead one imposes $\overline{{\rm tr}\,{\cal G}(K,K,A)}>0\,,\forall K$, where
\beq
\overline{{\cal O}(K,A)}\equiv \frac{1}{{\rm Vol}\,O(4)}\int_{\Lambda\in O(4)}{\cal O}(\Lambda K,A)\,.
\eeq
 This defines a priori a larger domain in $A$-space but again, when approaching the Gribov horizon from inside the Gribov region, at least one of the $\overline{{\rm tr}\,{\cal G}(K,K,A)}$'s has to change sign and the Gribov horizon lies inside the boundary of the region defined by $\overline{{\rm tr}\,{\cal G}(K,K,A)}>0\,,\forall K$. Let us also mention that, for the practical evaluation of  $\overline{{\rm tr}\,{\cal G}(K,K,A)}$, one can always assume that $A$ is transverse.

Given these preliminary remarks, at order $g^2$, one finds
\cite{Vandersickel:2011zc}
\begin{widetext}
\beq
{\cal G}_{ab}(K,P;A) & = & \frac{(2\pi)^d}{K^2}\delta_{ab}\delta^{(d)}(K-P)+igf_{bda}\frac{1}{K^2}\frac{P_\mu}{P^2}A^d_\mu(K-P)\nonumber\\
& + & (ig)^2f_{cda}f_{bec}\frac{1}{K^2}\int_Q\frac{(K-Q)_\mu}{(K-Q)^2}\frac{P_\nu}{P^2}A^d_\mu(Q)A^e_\nu(K-Q-P)\,,
\eeq
\end{widetext}
and therefore
\beq
\frac{1}{V_d d_G}{\rm tr}\,\overline{{\cal G}(K,K;A)}=\frac{1}{K^2}\Big[1+\sigma(K^2,A)\Big]\,,
\eeq
with
\beq
\sigma(K^2,A) & = & \frac{1}{V_d(d-1)}\frac{g^2C_{\rm ad}}{d_G}P^\parallel_{\mu\nu}(K)\nonumber\\
& & \times\,\int_Q\frac{P^\perp_{\mu\nu}(Q)}{(K-Q)^2}\overline{A^\alpha_\rho(Q)A^\alpha_\rho(-Q)}\,,
\eeq
where the labels $\alpha$ and $\rho$ are summed over. In deriving this expression, we have used that $A$ can be taken transverse, and, by using appropriate changes of variables, we have traded the average over $O(4)$ Euclidean rotations of $K$ by the average
\beq
\overline{A^\alpha_\rho(Q)A^\alpha_\rho(-Q)}= \frac{1}{{\rm Vol}\,O(4)}\int_{\Lambda\in O(4)}\!\!\!\!A^\alpha_\rho(\Lambda Q)A^\alpha_\rho(-\Lambda Q)\,.\nonumber\\
\eeq
The previous formula corresponds to the strict expansion of a propagator to order $g^2$. To this order, this is equivalent to
\beq
\frac{1}{V_d d_G}\overline{{\rm tr}\,{\cal G}(K,K,A)}=\frac{1}{K^2}\frac{1}{1-\sigma(K^2,A)}\,,
\eeq
In this 1PI-resummed form, the result is expected to be more accurate.

The Gribov no-pole condition corresponds a priori to the infinite set of conditions 
\beq\label{eq:G_full}
\forall K\,,1-\sigma(K^2,A)>0\,.
\eeq
However, it is usually argued that it is enough to impose the no-pole condition in the form
\beq\label{eq:G_0}
1-\sigma(0,A)>0\,.
\eeq 
This is because $\sigma(K^2,A)$ is a decreasing function of $K^2$. In fact, because $\overline{A^\alpha_\rho(Q)A^\alpha_\rho(-Q)}$ depends only on $Q^2$, the dependence with respect to $K$ originates only from the angular integral
\beq
\Omega_d(K^2/Q^2)\equiv \int_0^\pi  \frac{d\theta\,\sin^d\theta}{K^2/Q^2+1-2K/Q\cos\theta}\,.
\eeq
In Fig.~\ref{fig:angular}, we show that $\Omega_d(x)$ is a decreasing function of $x>0$, for $d\geq 2$ and, since $\overline{A^\alpha_\rho(Q)A^\alpha_\rho(-Q)}$ is positive, it follows that $\sigma(K^2,A)$ decreases indeed with $K^2$.\\

\begin{figure}[t]  
\epsfig{file=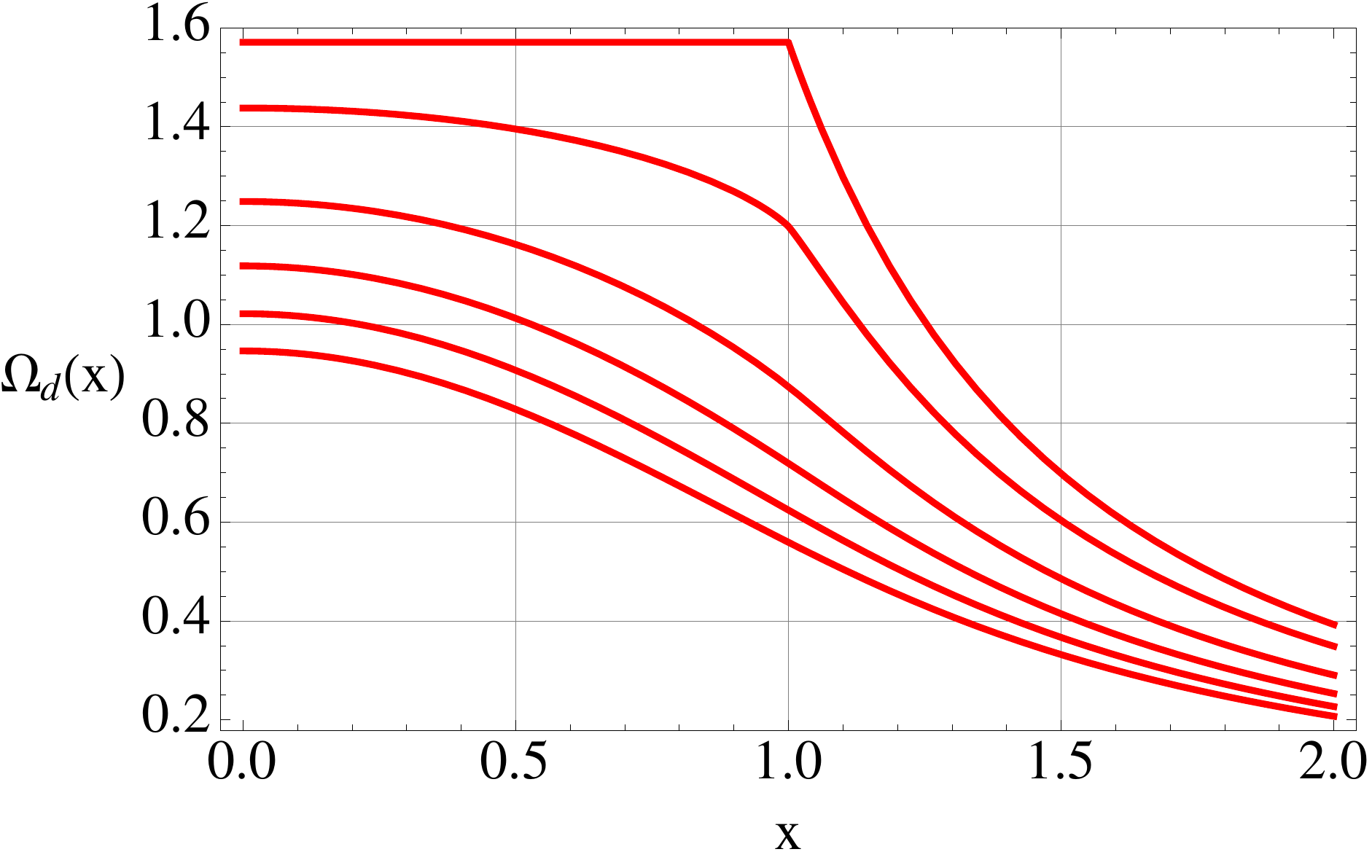,width=7.5cm}
\caption{The function $\Omega_d(x)$ for $d\geq 2$ and $x>0$.}\label{fig:angular}
\end{figure}

In the limit $K\to 0$, one finds
\beq
\sigma(0,A)=\frac{1}{V_dd}\frac{g^2C_{\rm ad}}{d_G}\int_Q\frac{A^\alpha_\rho(Q)A^\alpha_\rho(-Q)}{Q^2}\,,
\eeq
where we have used that $\int_Q\,f(Q^2)\,\overline{A^\alpha_\rho(Q)A^\alpha_\rho(Q)}=\int d^dQ\,f(Q^2)\,A^\alpha_\rho(Q)A^\alpha_\rho(Q)$. In order to implement the constraint (\ref{eq:G_0}), one then writes
\beq
& &\theta(1-\sigma(0,A))\propto\int_{-i\infty+\epsilon}^{+i\infty+\epsilon}\frac{d\beta}{2\pi i\beta}\,e^{\beta(1-\sigma(0,A))}\,.
\eeq
The partition function becomes
\beq\label{eq:toto7}
Z & = & \int{\cal D}Ac\bar ch\,\theta(1-\sigma(0,A))\,e^{-S_{\rm FP}[A,c,\bar c,h]}\nonumber\\
& = & \int_{-i\infty+\epsilon}^{+i\infty+\epsilon}\frac{d\beta}{2\pi i}\,\! e^{\beta-\ln\beta-V_d\,f(\beta)}.
\eeq
Given that, in the gluonic sector, the quadratic part of the action in Fourier space becomes (we introduce a gauge-fixing parameter $\xi$ that we will send to zero at the end)
\beq
K^{ab}_{\mu\nu}(Q)=\delta^{ab}\left[Q^2P^\perp_{\mu\nu}(Q)+\frac{Q^2}{\xi}P^\parallel_{\mu\nu}(Q)+\frac{m^4(\beta)}{Q^2}\delta_{\mu\nu}\right],\nonumber\\
\eeq
with
\beq\label{eq:defm}
m^4(\beta)\equiv \frac{2}{d}\frac{g^2C_{\rm ad}}{d_G}\frac{\beta}{V_d}\,,
\eeq
one obtains, at one-loop order
\beq
f(\beta) & = & d_G\left[\frac{d-1}{2}\int_Q\ln \frac{Q^4+m^4(\beta)}{Q^2}\right.\\
& & \hspace{0.6cm}+\,\left.\frac{1}{2}\int_Q\ln \frac{Q^4+\xi m^4(\beta)}{\xi Q^2}-\int_Q\ln Q^2\right]\,,\nonumber
\eeq
where the last term is the ghost contribution. One can evaluate the integral over $\beta$ using a saddle-point approximation. One finds $\ln Z\sim \beta_\star-\ln\beta_\star-V_d f(\beta_\star)$, with
\beq
0=1-\frac{1}{\beta_\star}-\frac{d-1}{d}g^2C_{\rm ad}\int_Q\frac{1}{Q^4+m^4(\beta_\star)}\,.
\eeq
If we assume $m^4(\beta_\star)$ to have a non-trivial infinite volume limit, $\beta_\star$ has to diverge linearly with $V_d$ and we arrive at a free-energy density that coincides with the zero-temperature and zero-background limit of Eq.~(\ref{eq:V}) with
\beq
0=1-\frac{d-1}{d}g^2C_{\rm ad}\int_Q\frac{1}{Q^4+m^4(\beta_\star)}\,.
\eeq

\vglue2mm

The extension to the Landau-DeWitt gauge is rather straightforward: one switches to a Cartan-Weyl basis (which implies in particular replacing $if^{abc}$ by $f^{\kappa\lambda\tau}$), replaces $A_\mu$ by $a_\mu$ and each momentum by its appropriately shifted version. One then considers the ghost propagator ${\cal G}_{\kappa\lambda}(K,P,a;\bar A)$ in the presence of a gauge-field configuration $a$ and a background $\bar A$, and evaluates
\beq\label{eq:82}
& & {\cal G}_{\kappa\kappa}(K_{-\kappa},K_{-\kappa},a;\bar A)\nonumber\\
& & \hspace{0.2cm}=\,\int_x\int_y (e^{iK_{-\kappa}x}\delta_{\kappa\eta})^*(-\bar D D)^{-1}_{\eta\xi}(x,y)(e^{iK_{-\kappa}y}\delta_{\kappa\xi})\,.\nonumber\\
\eeq
Again, if $a$ belongs to the first Gribov region, we have ${\cal G}_{\kappa\kappa}(K_{-\kappa},K_{-\kappa},a;\bar A)>0$, $\forall \kappa$ and $\forall K$. Similarly to the Landau gauge case, we shall impose instead $\overline{\sum_\kappa {\cal G}_{\kappa\kappa}(K_{-\kappa},K_{-\kappa},a;\bar A)}>0$,  $\forall K$, with
\beq
\overline{{\cal O}(K,a;\bar A)} & \equiv & \frac{1}{{\rm Vol}\,O(4)}\int_{\Lambda\in O(4)}{\cal O}(\Lambda K,a,\bar A)\,,
\eeq
and where we can assume that $a$ is transverse in a background covariant way. At order $g^2$, We find
\begin{widetext}
\beq
{\cal G}_{\kappa\lambda}(K,P,a;\bar A) & = & \frac{(2\pi)^d}{K^2_\kappa}\delta_{\kappa\lambda}\delta^{(d)}(K-P)+gf_{\lambda\eta\kappa}\frac{1}{K^2_\kappa}\frac{P^\lambda_\mu}{P_\lambda^2}a^\eta_\mu(K-P)\nonumber\\
& + & g^2f_{\eta \xi(-\kappa)}f_{\lambda \zeta(-\eta)}\frac{1}{K_\kappa^2}\int_Q\frac{(K_\kappa-Q_\xi)_\mu}{(K_\kappa-Q_\xi)^2}\frac{P^\lambda_\nu}{P^2_\lambda}a^\xi_\mu(Q)a^\zeta_\nu(K-Q-P)\,.
\eeq
\end{widetext}
and therefore
\beq
\frac{1}{V_d d_G}\overline{\sum_\kappa {\cal G}_{\kappa\kappa}(K_{-\kappa},K_{-\kappa},a;\bar A)}=\frac{1}{K^2}\frac{1}{1-\sigma(K^2,a;\bar A)},\nonumber\\
\eeq
with
\beq
\sigma(K^2,a;\bar A) & = & \frac{1}{V_{d}(d-1)}\frac{g^2C_{\rm ad}}{d_G}P^\parallel_{\mu\nu}(K)\\
& & \times\,\int_Q\frac{P^\perp_{\mu\nu}(Q)}{(K-Q)^2}\overline{a^\xi_\rho(Q_{-\xi}) a^{-\xi}_\rho(-Q_{-\xi})}\,.\nonumber
\eeq
In deriving these expressions, before taking the average over $\Lambda$-transformations,  we have used that, at zero temperature, one can always shift the integration momentum $Q_\xi$ to $Q$. Then, by appropriate changes of variables, we have traded the average over $O(4)$ Euclidean rotations of $K$ by the average
\beq
& & \overline{a^\xi_\rho(Q_{-\xi}) a^{-\xi}_\rho(-Q_{-\xi})}\\
& & \hspace{0.2cm}=\,\frac{1}{{\rm Vol}\,O(4)}\int_{\Lambda\in O(4)}\!\!\!\!a^\xi_\rho((\Lambda Q)_{-\xi}) a^{-\xi}_\rho(-(\Lambda Q)_{-\xi})\,.\nonumber
\eeq
It is easily checked that $\overline{a^\xi_\rho(Q_{-\xi}) a^{-\xi}_\rho(-Q_{-\xi})}$ is positive and depends only on $Q^2$. Therefore we are in a similar situation as above, with $\sigma(K^2,a;\bar A)<\sigma(0,a;\bar A)$ and
\beq
\sigma(0,a;\bar A) & = & \frac{1}{V_{d}d}\frac{g^2C_{\rm ad}}{d_G}\int_Q\frac{\overline{a^\xi_\rho(Q_{-\xi}) a^{-\xi}_\rho(-Q_{-\xi})}}{Q^2}\nonumber\\
& = & \frac{1}{V_{d}d}\frac{g^2C_{\rm ad}}{d_G}\int_Q\frac{a^\xi_\rho(Q) a^{-\xi}_\rho(-Q)}{Q^2_\xi}\,,
\eeq
where we made use of $\int_Q\,f(Q^2)\,\overline{a^\xi_\rho(Q_{-\xi}) a^{-\xi}_\rho(-Q_{-\xi})}=\int_Q\,f(Q^2)\,a^\xi_\rho(Q_{-\xi}) a^{-\xi}_\rho(-Q_{-\xi})$ and we changed the integration variable back to $Q_\xi$.

After introducing a parameter $\beta$ to impose the no-pole condition, we arrive at $\ln Z=\beta_\star-\ln\beta_\star-V_d \sum_\kappa f_\kappa(\beta_\star)$ with
\beq
f_\kappa(\beta) & = & \frac{d-1}{2}\int_Q\ln \frac{Q_\kappa^4+m^4(\beta)}{Q_\kappa^2}\nonumber\\
& + & \frac{1}{2}\int_Q\ln \frac{Q_\kappa^4+\xi m^4(\beta)}{\xi Q_\kappa^2}-\int_Q\ln Q^2_\kappa\,,
\eeq
and
\beq
m^4(\beta)=\frac{2}{d}\frac{g^2C_{\rm ad}}{d_G}\frac{\beta}{V_d}\,.
\eeq
The parameter $m^4$ is fixed through the saddle-point equation
\beq
1=\frac{d-1}{d}\frac{g^2C_{\rm ad}}{d_G}\sum_\kappa\int_Q\frac{1}{Q_\kappa^4+m^4}\,.
\eeq
This is nothing but the gap equation obtained with the model (\ref{eq:action_prop_upgrade_2}). Of course at zero-temperature, one can always shift the momenta $Q_\kappa$ back to $Q$, in which case the free-energy density and the gap equations coincide trivially with the ones obtained in the Landau gauge.

%%%
\subsection{Mapping to the Landau gauge}\label{sec:mappin_Landau}

The previous results are not surprising because, at zero temperature, the expression for the partition function in the Landau-DeWitt gauge can be related to the one in the Landau gauge through a trivial transformation of the fields, namely\footnote{Of course this does not mean that the two gauges are identical because the correlations functions are not the same. However they are related by trivial identities, see for instance \cite{Reinosa:2015gxn}.}
\beq\label{eq:map}
(X^U)^\kappa(x)=e^{i\tau g\bar A\cdot\kappa}X^\kappa(x)
\eeq
First, using the property
\beq
\bar D^\kappa_\mu (X^U)^\kappa(x)=e^{i\tau g\bar A\cdot\kappa}\partial_\mu X^\kappa(x)\,,
\eeq
\vglue1mm
\noindent{it is easily checked that, upon this change of variables, the Faddev-Popov action for the Landau-DeWitt gauge becomes the Faddeev-Popov action for the Landau gauge, after one renames $a_\mu$ into $A_\mu$. It is then easily checked that if one starts from the Gribov-Zwanziger action for the Landau gauge and apply the change of variables (after renaming $A_\mu$ into $a_\mu$)}
\beq\label{eq:map2.1}
(X^U)^\kappa(x) & = & e^{-i\tau g\bar A\cdot\kappa}X^\kappa(x)\,,\\
(X^U)^{\kappa\lambda}(x) & = & e^{-i\tau g\bar A\cdot(\kappa+\lambda)}X^{\kappa\lambda}(x)\,, \label{eq:map2.2}
\eeq
one obtains the action (\ref{eq:action_prop_upgrade}). 

We should mention however that this mapping crucially relies on the fact that the boundary conditions are not important at zero temperature, at least in the Faddeev-Popov framework. To check this, consider Yang-Mills fields on a compact time interval of length $L$ (which will eventually be sent to $\infty$) with boundary conditions of the form
\beq\label{eq:boundary}
{\rm bc1\!:} \,\,\, a^\kappa_\mu(\tau+L,\vec{x})=e^{ig\bar B\cdot\kappa}a^\kappa_\mu(\tau,\vec{x})\,,
\eeq
with $\bar B$ a constant vector in a space isomorphic to the Cartan subalgebra. For the partition function to be invariant under gauge transformations, the latter should be chosen to preserve the boundary (\ref{eq:boundary}). This means that the Faddeev-Popov procedure applied to the Landau-DeWitt gauge leads to the usual action but with the peculiarity that  all fields obey the boundary conditions (\ref{eq:boundary}).\footnote{Under an infinitesimal transformation we have $\delta a^\kappa_\mu=\partial_\mu\theta^\kappa -igf^{-\kappa\lambda\eta}\theta^\lambda a_\mu^\eta$. If $\theta^\kappa$ obeys the boundary conditions (\ref{eq:boundary}), then, using that $f^{-\kappa\lambda\eta}$ is color conserving, one finds that $\delta a^\kappa_\mu$ also obeys the boundary conditions (\ref{eq:boundary}).}

Consider now a two-point function (this could be any correlation function, including the partition function) ${\cal G}^{L,{\rm bc1}}_{\kappa\lambda}(x,y;\bar A)$, computed within this particular gauge-fixing. We will now show that, in the ``zero temperature'' limit ($L\to\infty$) it coincides with the same correlation function computed within the same gauge, but with periodic boundary conditions
\beq\label{eq:boundary2}
{\rm bc2\!:} \,\,\, a^\kappa_\mu(\tau+(L\to\infty),\vec{x})=a^\kappa_\mu(\tau,\vec{x})\,,
\eeq
To show this, we first apply the change of variables in \eqref{eq:map2.1} and \eqref{eq:map2.2} with $\bar A$ replaced by $\bar B$. This turns the boundary conditions of all fields into periodic ones, while changing the background from $\bar A$ to $\bar A+\bar B$ and multiplying all correlation functions by appropriate phase factors:
\beq
{\cal G}^{L,{\rm bc1}}_{\kappa\lambda}(x,y;\bar A)=e^{-i(\tau \kappa+\tau' \lambda)\cdot g\bar B}{\cal G}^{L,{\rm bc2}}_{\kappa\lambda}(x,y;\bar A+\bar B)\,.\nonumber\\
\eeq
Next, one applies a background gauge transformations to obtain
\beq
{\cal G}^{L,{\rm bc1}}_{\kappa\lambda}(x,y;\bar A)=e^{-i(\tau \kappa+\tau' \lambda)\cdot g\bar B'}{\cal G}^{L,{\rm bc2}}_{\kappa\lambda}(x,y;\bar A+\bar B')\,,\nonumber\\
\eeq
with $\bar B'=\bar B-n\bar\alpha/(gL)$, for any $\bar\alpha$ that maintains the $1/L$-periodicity of the fields and for any $n\in\mathds{Z}$. Taking the ``zero temperature'' limit as $L=n/(ug)$ with $u$ any real number and $n\to\infty$, we arrive at
\beq
{\cal G}^{\infty,{\rm bc1}}_{\kappa\lambda}(x,y;\bar A)=e^{-i(\tau \kappa+\tau' \lambda)\cdot g\bar B'}{\cal G}^{\infty,{\rm bc2}}_{\kappa\lambda}(x,y;\bar A+\bar B')\,,\nonumber\\
\eeq
with $\bar B'=\bar B-u\bar\alpha$. Since the $\bar\alpha$'s form a basis of the Cartan subalgebra, repeated use of the previous formula leads to
\beq
{\cal G}^{\infty,{\rm bc1}}_{\kappa\lambda}(x,y;\bar A)={\cal G}^{\infty,{\rm bc2}}_{\kappa\lambda}(x,y;\bar A)\,.
\eeq
As announced, the zero-temperature correlations functions in the Faddeev-Popov gauge-fixing are the same for the two sets of boundary conditions.

It is however not clear how these remarks extend to the Gribov gauge-fixing. In particular, we should notice that in the above derivation, the use of shifted momenta in (\ref{eq:82}) implicitly restricts the search for eigenstates of the Faddeev-Popov operator to eigenstates with certain boundary conditions, those that are precisely mapped to the periodic eigenstates in the Landau gauge. It is not clear to us whether this is what should be done or how taking into account other boundary conditions would affect the result.

%%%
\subsection{Extension to finite temperature?}\label{appsec:GribovT}

The problem with the boundary conditions is even more visible at finite temperature. First of all, in this case, there is no change of variables that allows to get rid of the background, since the allowed transformations are constrained by the periodicity of the fields. Moreover, as it has been discussed in Refs.~\cite{Cooper:2015sza,Cooper:2015bia}, the periodic boundary conditions directly affect the implementation of the Gribov gauge-fixing via the Gribov-Zwanziger construction.\footnote{We shall not discuss it here but the implementation of the Gribov no-pole consition is also substantially modified.} Let us here summarize the argument in the case of the Landau gauge and then briefly speculate on the consequences for the Landau-DeWitt gauge. A more detailed discussion is postponed for a future investigation.

The Gribov-Zwanziger construction is based on the perturbative evaluation of the lowest non-zero eigenvalues of the Faddeev-Popov operator, starting from the lowest non-zero (degenerate) eigenvalue of the free Faddeev-Popov operator. At zero temperature, working in a box of volume $L^4$ with periodic boundary conditions, the eigenstates of the free Faddeev-Popov operator are of the form $\exp(i(2\pi/L) (n_0\tau+\vec{n}\cdot\vec{x}))$, with $n_\mu\in\mathds{Z},\,\forall \mu$, and the corresponding eigenvalues are $(2\pi/L)^2 (n_0^2+||\vec{n}||^2)$. Therefore, the lowest non-zero eigenvalue corresponds to states with $\smash{n_0^2+||\vec{n}||^2=1}$. In contrast, at finite temperature, where the system is in a box of size $\beta L^3$, the periodic eigenstates are rather $\exp(i(2\pi/\beta) n_0\tau +(2\pi/L \vec{n})\cdot\vec{x})$ and the corresponding eigenvalues are $(2\pi/\beta)^2 n_0^2+(2\pi/L)^2||\vec{n}||^2$. Therefore, in this case, the smallest, non-zero eigenvalue corresponds to states with $\smash{n_0=0}$ and $\smash{||\vec{n}||^2=1}$. This has a direct imprint on the Gribov-Zwanziger construction and leads to an action that is not simply the zero-temperature Gribov-Zwanziger action taken over a compact time interval, see Refs.~\cite{Cooper:2015sza,Cooper:2015bia} for more details.

We mention here that, even though this asymmetrical treatment of the temporal and spatial components is to be expected at finite temperature, it leads to some unexpected features. In particular, in the zero-temperature limit, one does not recover the usual Gribov-Zwanziger action but rather an action that explicitly breaks the Euclidean $O(4)$ invariance of the vacuum theory. This raises some conceptual issues, in particular concerning the renormalizability of the action or the potential contamination of the zero-temperature observables by these $O(4)$-breaking terms. Of course, if the Gribov-Zwanziger construction corresponds to a bona-fide gauge-fixing, we expect the $O(4)$ breaking terms to be restricted to the gauge-fixing sector and not to affect the $O(4)$-invariance or the UV finiteness of the zero-temperature observables. However, since the Gribov restriction is never implemented exactly in practice,\footnote{In the case of the Gribov-Zwanziger approach, even though the true condition should be that the smallest value of the Faddeev-Popov operator to remain positive, in practice, one imposes the sum of the smallest eigenvalues (as described above) to remain positive, which obviously does not imply that the smallest one is positive. In fact, in our view this is the reason why the Gribov-Zwanziger approach and the zero-temperature limit do not seem to commute.} these issues deserve a careful investigation.

We leave this interesting questions for a future work and end this section by speculating on the implications of the previous remarks for the Landau-DeWitt gauge. In the Landau-DeWitt gauge at finite temperature, the r\^ole of the free Faddeev-Popov operator is played by $\bar D^2$ but the fields remain periodic. Therefore, the eigenstates are still of the form $\exp(i(2\pi/\beta)n_0\tau+(2\pi/L)\vec{n}\cdot\vec{x})$ but the eigenvalues become $(2\pi n_0+r\cdot\kappa)^2/\beta^2+(2\pi/L)^2||\vec{n}||^2$. It follows that, for generic backgrounds such that $r\cdot\kappa$ is not a multiple of $2\pi$, the lowest non-zero eigenvalues correspond to $\smash{\kappa=0}$, $\smash{n_0=0}$ and $\smash{||\vec{n}||^2=1}$. So not only would the Gribov-Zwanziger procedure affect only the spatial components of the gauge field but only those color components that are aligned with the background. In this case the order parameter for the deconfinement transition -- the Polyakov loop or the background $\bar A$ at the minimum of the background effective potential -- would not interact with the Gribov region at one-loop order, in contrast to what happens in the present work or in \cite{Canfora:2015yia}; the search for possible effects on the deconfinement transition would necessarily start at two-loop order.

\section{Conclusions and Outlook}
We have put forward a Gribov-Zwanziger type action for the Landau-DeWitt gauge that remains invariant under background gauge transformations. At zero-temperature and to one-loop accuracy, our model can be related to the Gribov no-pole condition applied to the Landau-DeWitt gauge. Moreover, in contrast to other recent proposals, our model does not require the introduction of a Stueckelberg field.

Without spoiling the background gauge invariance, our approach allows for color dependent Gribov parameters, a possibility which we have investigated together with its impact on the deconfinement transition. We have observed variations of the transition temperature up to $20\%$. We have also observed that certain Gribov parameters can become negative while maintaining a real effective potential. In fact, in some cases, the transition is only properly accounted for if $m^4$ is allowed to become negative. We mention that, in a recent study, the three scenarios proposed here have been tested against lattice simulations \cite{Maelger:2018vow}. The degenerate scenario seems to be favoured.

Our model allows for the evaluation of higher corrections in a manifestly background gauge invariant way. We are currently evaluating the two-loop background effective potential and the corresponding finite temperature two-loop gap equations for the Gribov parameters.

Finally, it is important to mention that, at finite temperature, none of the existing proposals, including ours, can be understood so far as faithful implementations of the Gribov-Zwanziger restriction for the Landau-DeWitt gauge. In this respect, it would be important to generalize the considerations of Refs.~\cite{Cooper:2015sza,Cooper:2015bia} to the Landau-DeWitt gauge, along the lines of the discussion that we have initiated in Sec.~\ref{appsec:GribovT}.

%%%%%
\acknowledgements{We would like to thank David Dudal and C\'edric Lorc\'e for useful discussions. D.~K. acknowledges support from the ``Physique des 2 Infinis et des Origines'' Labex (P2IO).}

\appendix

%%%%%
\section{Path (in-)dependence}\label{app:Wilson}
Even though we did not make it explicit in our notation, for general backgrounds (such that $\bar F_{\mu\nu}\neq 0$) the redefinitions $\hat\varphi_\nu(x)$ and $\hat\omega_\nu(x)$ of the fields $\varphi_\nu(x)$ and $\omega_\nu(x)$ through the Wilson line (\ref{eq:Wilson}) are not true functions of $x$, since they also depend on the path $C$ used to define the Wilson line. Therefore, we need to be more specific about what is meant by ``covariant derivatives'' acting on this type of objects in Eq.~(\ref{eq:action_prop_mod}). We write our action proposal as
\beq\label{eq:S_Delta}
S_{\rm new} & = & \int_x \bigg\{\frac{1}{4}F_{\mu\nu}^a F_{\mu\nu}^a+ih^a \bar D^{ab}_\mu a^b_\mu+\bar c^a \bar D_\mu^{ab} D^{bc}_\mu c^c\nonumber\\
& & \hspace{0.7cm}-\, (\hat\omega^\dagger_\nu)^{ea} \bar \Delta^{ab}_\mu \Delta^{bc}_\mu \hat\omega^{ce}_\nu+(\hat\varphi^\dagger_\nu)^{ea} \bar \Delta^{ab}_\mu \Delta^{bc}_\mu\hat\varphi^{ce}_\nu\nonumber\\
& & \hspace{0.7cm}-\,g \gamma^{1/2} f^{abc} a_\mu^a(\varphi^{bc}_\mu+\bar\varphi^{bc}_\mu)-\gamma dd_G\bigg\}\,,
\eeq
and define
\beq\label{eq:Delta}
\Delta_\mu\hat\varphi_\nu(x) & \equiv & (D_\mu\varphi_\nu(x))L_{\bar A,C}(x,x_0)\nonumber\\
& & +\,ig\varphi_\nu(x)\bar A_\mu^a(x)T^aL_{\bar A,C}(x,x_0)\,,
\eeq
and similarly for $\bar\Delta_\mu$. These definitions coincide with the usual covariant derivatives in the case where $\bar F_{\mu\nu}=0$. Moreover, by noticing that the RHS of (\ref{eq:Delta}) is a linear combination of true functions multiplied by the Wilson line, the repeated action of such operators can be simply defined by assuming that $\Delta_\mu$ acts linearly on this type of linear combinations.

The definition (\ref{eq:Delta}) is similar in spirit to the so-called Mandelstam derivative of the Wilson line \cite{}. We stress however that, because it does not apply to functions (unless $\bar F_{\mu\nu}=0$), this is not a true derivative and thus it should not be used as such (a similar word of caution applies to the Mandelstam derivative). To make this point clear, we use the notation $\Delta_\mu$ for the rest of this section. In the manipulations to be discussed now, we shall always rely on the above definition and will not assume without proof that $\Delta_\mu$ shares the same properties as a derivative operator. For instance, it will be convenient to show that given two objects $\hat\varphi$ and $\hat\psi$ of the form ``function times Wilson line'', the following formula of integration by parts holds:
\beq\label{eq:ibp}
\int_x {\rm tr}\,\hat\psi^\dagger \Delta_\mu\hat\varphi=-\int_x {\rm tr}\,(\Delta_\mu\hat\psi)^\dagger \hat\varphi\,.
\eeq
To this purpose, we write
\beq
& & {\rm tr}\,(\psi L_{\bar A,C})^\dagger \Delta_\mu(\varphi L_{\bar A,C})+{\rm tr}\,(\Delta_\mu(\psi L_{\bar A,C}))^\dagger \varphi L_{\bar A,C}\nonumber\\
& & = {\rm tr}\,L_{\bar A,C}^\dagger\psi^\dagger (D_\mu\varphi) L_{\bar A,C}+{\rm tr}\,L_{\bar A,C}^\dagger(D_\mu \psi^\dagger) \varphi L_{\bar A,C}\nonumber\\
& & + ig\,{\rm tr}\,L_{\bar A,C}^\dagger \psi^\dagger\varphi\bar A_\mu^aT^aL_{\bar A,C}-ig\,{\rm tr}\,L_{\bar A,C}^\dagger \bar A_\mu^aT^a\psi^\dagger\varphi L_{\bar A,C}\nonumber\\
& & = {\rm tr}\,\partial_\mu(\psi^\dagger\varphi)-g\,(\psi^\dagger)^{ea}f^{abc}A_\mu^c\varphi^{be}-g\,f^{eac}A_\mu^c(\psi^\dagger)^{ab}\varphi^{be}\nonumber\\
& & ={\rm tr}\,\partial_\mu(\psi^\dagger\varphi)\,.
\eeq
In the last step, we have used that the fields $\varphi^{ab}$ and $\psi^{ab}$ are antisymmetric. An integration over $x$ leads finally to (\ref{eq:ibp}).

We are now ready to check the background gauge-inviance of (\ref{eq:S_Delta}) in a more rigorous way. To that aim, we first use the integration by parts formula (\ref{eq:ibp}) to rewrite the action as
\beq\label{eq:S_Delta_2}
S_{\rm new} & = & \int_x \bigg\{\frac{1}{4}F_{\mu\nu}^a F_{\mu\nu}^a+ih^a \bar D^{ab}_\mu a^b_\mu+\bar c^a \bar D_\mu^{ab} D^{bc}_\mu c^c\nonumber\\
& & \hspace{0.7cm}+\,{\rm tr}\,(\bar\Delta_\mu\hat\omega_\mu)^\dagger  (\Delta_\mu \hat\omega_\nu)-{\rm tr}\,(\bar\Delta\hat\varphi_\nu)^\dagger (\Delta_\mu\hat\varphi_\nu)\nonumber\\
& & \hspace{0.7cm}-\,g \gamma^{1/2} f^{abc} a_\mu^a(\varphi^{bc}_\mu+\bar\varphi^{bc}_\mu)-\gamma dd_G\bigg\}.
\eeq
Then, we evaluate
\beq
& & \Delta^U_\mu(\varphi^U_\nu(x)L_{\bar A^U,C}(x,x_0))\nonumber\\
& & \hspace{0.5cm}\equiv\,(D^U_\mu\varphi^U_\nu(x))L_{\bar A^U,C}(x,x_0)\nonumber\\
& & \hspace{0.5cm}+\,ig\varphi^U_\nu(x)(A^U_\mu)^a(x)T^aL_{\bar A^U,C}(x,x_0)\nonumber\\
& & \hspace{0.5cm}=\,D^U_\mu({\cal U}(x)\varphi_\nu(x))L_{\bar A,C}(x,x_0){\cal U}^\dagger(x_0)\nonumber\\
& & \hspace{0.5cm}+\,{\cal U}(x)\varphi_\nu(x)\partial_\mu{\cal U}^\dagger(x){\cal U}(x)L_{\bar A,C}(x,x_0){\cal U}^\dagger(x_0)\nonumber\\
& & \hspace{0.5cm}+\,ig\,{\cal U}(x)\varphi_\nu(x)A_\mu^a(x)T^aL_{\bar A,C}(x,x_0){\cal U}^\dagger(x_0)\nonumber\\
& & \hspace{0.5cm}-\,{\cal U}(x)\varphi_\nu(x)\partial_\mu{\cal U}^\dagger(x){\cal U}(x) L_{\bar A,C}(x,x_0){\cal U}^\dagger(x_0)\nonumber\\
& & \hspace{0.5cm}=\,{\cal U}(x)(D_\mu\varphi_\nu(x))L_{\bar A^U,C}(x,x_0){\cal U}^\dagger(x_0)\nonumber\\
& & \hspace{0.5cm}+\,ig\,{\cal U}(x)\varphi_\nu(x)A_\mu^a(x)T^aL_{\bar A,C}(x,x_0){\cal U}^\dagger(x_0)\nonumber\\
& & \hspace{0.5cm}={\cal U}(x)\left[\Delta_\mu(\varphi_\nu(x)L_{\bar A,C}(x,x_0))\right]{\cal U}^\dagger(x_0)\,.
\eeq
The background gauge invariance of (\ref{eq:S_Delta_2}) and therefore of (\ref{eq:S_Delta}) follows immendiately. 

We can also check the independence of our procedure with respect to the chosen path. Indeed, if we consider a second path $C'$, we have
\beq
\hat\varphi'(x) & = & \varphi(x)L_{\bar A,C'}(x,x_0)\nonumber\\
& = & \hat\varphi(x)L^{-1}_{\bar A,C}(x,x_0)L_{\bar A,C'}(x,x_0)\,.
\eeq
By definition
\beq
\Delta_\mu\hat\varphi'(x) & \equiv & (D_\mu\varphi(x))L_{\bar A,C'}(x,x_0)\nonumber\\
& & +\,ig\varphi(x)A_\mu^a(x)T^a L_{\bar A,C'}(x,x_0)\,.
\eeq
Using (\ref{eq:Delta}), it is trivially seen that
\beq
\Delta_\mu\hat\varphi'(x)=(\Delta_\mu\hat\varphi(x))L^{-1}_{\bar A,C}(x,x_0)L_{\bar A,C'}(x,x_0)\,.
\eeq
Therefore, the second line of (\ref{eq:S_Delta_2}) can be reexpressed identically in terms of $\hat\varphi'$ and $\hat\omega'$. This completes the proof that our procedure is independent of the chosen path $C$, as announced above.

%%%%%
\section{Change to a Cartan-Weyl basis}\label{appsec:basis}
The change from a Cartersian basis $\{it^a\}$ to a Cartan-Weyl basis $\{it^\kappa\}$ is a change of basis in the complexified version of the Lie algebra. Therefore, in what follows, it will be convenient to introduce a formal complex conjugation to distinguish the elements of the original (real) Lie algebra, such that $\smash{\bar X=X}$, from those in the purely imaginary component of the complexified algebra, such that $\smash{\bar X=-X}$. In the case of $SU(N)$, where the elements of the original Lie algebra are antihermitian matrices, this complex conjugation can be represented as $\smash{\bar X\equiv-X^\dagger}$.\footnote{This formal complex conjugation should not be mistaken, however, with the standard complex conjugation of matrices, which we denote by $X^*$.} In particular, we have $\smash{\overline{it^a}=it^a}$. The Cartan-Weyl basis can always be chosen such that $\overline{it^\kappa}=it^{-\kappa}$. In particular, if
\beq
X=X_a\,it_a=X_\kappa\,it_\kappa\,,
\eeq
then
\beq
\bar X=(X_a)^*\,it_a=(X_{-\kappa})^*\,it_\kappa\,.
\eeq
This is exactly as with the Fourier transformation, for which $\smash{\bar X(Q)=(X(-Q))^*}$. If the field is real (meaning $\smash{\bar X=X}$), we find of course $(X_a)^*=X_a$ and $(X_{-\kappa})^*=X_\kappa$.\\

The change to a Cartan-Weyl basis is in fact an orthonormal change of basis if we equip the complexified algebra with the hermitian product
\beq
\langle X;Y\rangle=2\,{\rm tr}\,X^\dagger Y=-2\,{\rm tr}\,\bar XY\,.
\eeq
It follows that
\beq
(X_a)^* Y_a= (X_\kappa)^* Y_\kappa\,,
\eeq
which also rewrites
\beq
\bar X_a Y_a=\bar X_{-\kappa} Y_\kappa\,.
\eeq
This is similar to the Parseval-Plancherel identity. This identity has been extensively used in deriving Eq.~(\ref{eq:action_prop_upgrade}). 

We mention finally that in Eq.~(\ref{eq:action_prop_upgrade}), the components $\varphi^{\kappa\xi}_\nu$ or $\hat\varphi^{\kappa\xi}_\nu$ are tensor components whereas in Eq.~(\ref{eq:mat}) the same notation stands for matrix components. The reason why we use tensor components in Eq.~(\ref{eq:action_prop_upgrade}) is that the derivation is simpler for it relies directly on the identities given above. The matrix notation was useful in Sec.~\ref{sec:IIB} to identify invariant terms in the action. Changing from the matrix components to the tensor components simply amounts to changing the sign of $\xi$. To see this let us write the unitary change from Cartesian to Cartan-Weyl coordinates as
\beq
X_\kappa=M_{\kappa a} X_a\,,
\eeq
with $M^\dagger=M^{-1}$. Since this change of variables applies to any element of the complexified algebra, in particular to those in the real part, we have
\beq
X_{-\kappa}=M_{\kappa a}^* X_a\,,
\eeq
from which we deduce that $\smash{M_{\kappa a}^*=M_{(-\kappa)a}}$ and then $M^{-1}_{b\xi}=M^\dagger_{b\xi}=M^*_{\xi b}=M_{(-\xi)b}$. Let us now write the matrix and tensor components of $\varphi$ in the Cartan-Weyl basis respectively as
\beq
\tilde\varphi^{\kappa\xi} & = & M_{\kappa a}\varphi^{ab} M^{-1}_{b\xi}\,,\\
\varphi^{\kappa\xi} & = & M_{\kappa a}M_{\xi b}\varphi^{ab}\,.
\eeq
It follows that
\beq
\tilde\varphi^{\kappa\xi}=M_{\kappa a}M_{(-\xi) b}\varphi_{ab} =\varphi_{\kappa(-\xi)}\,,
\eeq
as announced.

%%%%%
\section{Gaussian integrals}\label{appsec:gaussian}
In some cases, we need to evaluate Gaussian integrals that mix real and complex variables. 
Consider for instance the integral
\beq
I=\int d^nx\,\frac{d^nz d^nz^*}{i}\,e^{-\frac{1}{2}X^{\rm t}MX-Z^\dagger N Z-X^{\rm t}P^\dagger Z-Z^\dagger PX}\,,\nonumber\\
\eeq
with $M$ real and symmetric and $N$ hermitian, both positive definite, so that the ``action'' is real and the integral absolutely convergent. We can integrate over the complex variables first. Using the change of variables
\beq
Z & \to & Z-N^{-1}PX\,,\\
Z^* & \to & Z^*-(N^*)^{-1} P^* X\,,
\eeq 
we find
\beq
I=\int d^nx\,\frac{d^nz d^nz^*}{i}\,e^{-\frac{1}{2}X^{\rm t}(M-2P^\dagger N^{-1}P)X-Z^\dagger N Z}\,.\nonumber\\
\eeq
After symmetrization of the newly obtained real quadratic form, we find
\beq\label{eq:result}
I=\frac{(2\pi)^{3n/2}}{{\rm det}\,N\,\sqrt{{\rm det}\,(M-P^\dagger N^{-1}P-(P^\dagger N^{-1}P)^{\rm t})}}\,.\nonumber\\
\eeq
A mnemonic way to recall this result is to rewrite the original integral as
\beq\label{eq:Ire}
I=\int d^{3n}\xi\,e^{-\frac{1}{2}\chi^\dagger {\cal N} \chi}\,,
\eeq
with
\beq
\chi=\left(
\begin{array}{c}
X\\
Z\\
Z^*
\end{array}
\right)
\eeq
and
\beq
{\cal N}=\left(
\begin{array}{ccc}
M & P^\dagger & P^{\rm t}\\
P & N & 0\\
P^* & 0 & N^{\rm t}
\end{array}
\right).
\eeq
A simple calculation, using Schur decomposition leads to 
\beq
{\rm det}\,{\cal N} & = & {\rm det} N \times {\rm det}\,\left(\begin{array}{cc}
M-P^{\rm t} (N^{\rm t})^{-1}P^* & P^\dagger\\
P & N
\end{array}\right)\nonumber\\
& = & ({\rm det}\,N)^2\times {\rm det}\,(M-P^\dagger N^{-1}P-(P^\dagger N^{-1}P)^{\rm t})\,.\nonumber\\
\eeq
Therefore, we can rewrite the result (\ref{eq:result}) as 
\beq
I=\frac{(2\pi)^{3n/2}}{\sqrt{{\rm det}\,{\cal N}}}\,,
\eeq
that is as it would result from (\ref{eq:Ire}) by considering the integral as a purely real one (i.e. disregarding the presence of the dagger and the fact that some of the components of $\chi$ are complex).

This is the reason why we have written the quadratic part of the action (\ref{eq:action_prop_upgrade_3}) in the $A-\varphi-\bar\varphi$ sector as 
\beq
\frac{1}{2}\int_{x,y}\chi^\dagger(x)\,{\cal M}(x-y)\,\chi(y)\,,
\eeq 
with $\chi^\dagger(x)=(a^\kappa_\mu(x),h^\lambda(x),\varphi^{\eta\xi}_\rho(x),\bar\varphi^{\bar\eta\bar\xi}_{\bar\rho}(x))^*$. This vector contains real components $a_\mu^{0^{(j)}}(x)$ and $h^{0^{(j)}}(x)$, as well as complex conjugated components $a_\mu^{\alpha}(x)$ and $a_\mu^{-\alpha}(x)$, $h^\alpha(x)$ and $h^{-\alpha}(x)$, $\varphi^{\eta\xi}_\rho(x)$ and $\bar\varphi^{(-\eta)(-\xi)}_\rho(x)$, and finally $\bar\varphi^{\bar\eta\bar\xi}_{\bar\rho}(x)$ and $\varphi^{(-\bar\eta)(-\bar\xi)}_{\bar\rho}(x)$.

\begin{widetext}

%%%%%
\section{Formulae}\label{appsec:formulae}
In what follows, we derive various formulae used in the main text. It will be important to allow for negative values of the Gribov parameter $m^4$ in those sum-integrals where the frequency is shifted by $r\cdot\kappa$. In fact the parameter $m^4$ can take values down to $-M_{r\cdot\kappa}^4$ with $M_{r\cdot\kappa}^4\equiv{\rm min}_{n\in\mathds{Z}}(2\pi n+r\cdot\kappa)^4T^4$.

%%%
\subsection{Sum-integral entering the gap equation}
The gap equation involves the sum-integral 
\beq
\hat J_\kappa(m^4)\equiv\int_Q^T \frac{1}{Q^4_\kappa+m^4}\,.
\eeq 
At zero temperature, it does not depend on the background since the latter can be shifted away by a change of variables. In that case, the Gribov parameter $m^4$ should be taken positive (without loss of generality, we can assume that $m^2>0$). We can then use
\beq\label{eq:decomp}
\frac{1}{Q^4+m^4}=-\frac{1}{m^2}{\rm Im}\,\frac{1}{Q^2+im^2}\,,
\eeq
together with the formula
\beq
\int_Q \frac{1}{Q^2+M^2}=-\frac{M^2}{16\pi^2}\left[\frac{1}{\epsilon}+\ln\frac{\bar\mu^2}{M^2}+1\right],
\eeq
valid for any non-negative (possibly complex) $M^2$, to arrive at
\beq\label{eq:toto3}
\hat J(m^4)\equiv\int_Q \frac{1}{Q^4+m^4}=\frac{1}{16\pi^2}\left[\frac{1}{\epsilon}+\frac{1}{2}\ln\frac{\bar\mu^4}{m^4}+1\right].
\eeq
\vglue4mm

We can proceed similarly at finite temperature, but this time we need to distinguish the cases $m^4>0$ and $-M_{r\cdot\kappa}^4<m^4<0$. If $m^4>0$, we use again (\ref{eq:decomp}) and the usual formula for the tadpole sum-integral at finite temperature. We find
\beq\label{eq:u}
\hat J_\kappa(m^4) & = & -\frac{1}{m^2}\int_q {\rm Im}\,\left[\frac{1+n_{\sqrt{q^2+im^2}-iT r\cdot\kappa}+n_{\sqrt{q^2+im^2}+iT r\cdot\kappa}}{2\sqrt{q^2+im^2}}\right].
\eeq
Because $m^2$ is real, the contribution $1$ in the numerator leads to the zero temperature limit (\ref{eq:toto3}). Rewriting also the finite temperature contribution in a simpler way, we arrive at
\beq\label{eq:toto2}
\Delta \hat J_\kappa(m^4;m^4_{\rm vac}) & \equiv & \hat J_\kappa(m^4)-\hat J_{T=0}(m^4_{\rm vac})\nonumber\\
& = & \frac{1}{32\pi^2}\ln\frac{m^4_{\rm vac}}{m^4}-\frac{1}{m^2}\int_q {\rm Im}\frac{1}{\sqrt{q^2+im^2}}\frac{e^{\sqrt{q^2+im^2}/T}\cos(r\cdot\kappa)-1}{e^{2\sqrt{q^2+im^2}/T}-2e^{\sqrt{q^2+im^2}/T}\cos(r\cdot\kappa)+1}\,,
\eeq
which we also rewrite for later convenience as
\beq\label{eq:u}
\Delta \hat J_\kappa(m^4;m^4_{\rm vac})=\frac{1}{32\pi^2}\ln\frac{m^4_{\rm vac}}{m^4} & + & \frac{1}{2im^2}\int_q \frac{1}{2\sqrt{q^2+im^2}}\frac{\cos(r\cdot\kappa)-e^{-\sqrt{q^2+im^2}/T}}{\cos(r\cdot\kappa)-\cosh(\sqrt{q^2+im^2}/T)}\nonumber\\
& - & \frac{1}{2im^2}\int_q \frac{1}{2\sqrt{q^2-im^2}}\frac{\cos(r\cdot\kappa)-e^{-\sqrt{q^2-im^2}/T}}{\cos(r\cdot\kappa)-\cosh(\sqrt{q^2-im^2}/T)}\,.
\eeq
If $-M_{r\cdot\kappa}^4<m^4<0$, we write $m^2=iM^2$ (we can assume that $M^2>0$) and use again (\ref{eq:decomp}) but rather as a difference. We find
\beq\label{eq:eq}
\hat J_\kappa(m^4) & = & \frac{1}{2M^2}\int_{q<M} \frac{1+n_{i\sqrt{M^2-q^2}-iT r\cdot\kappa}+n_{i\sqrt{M^2-q^2}+iT r\cdot\kappa}}{2i\sqrt{M^2-q^2}}\nonumber\\
& & +\frac{1}{2M^2}\int_{q>M} \frac{1+n_{\sqrt{q^2-M^2}-iT r\cdot\kappa}+n_{\sqrt{q^2-M^2}+iT r\cdot\kappa}}{2\sqrt{q^2-M^2}}\nonumber\\
& & -\frac{1}{2M^2}\int_q\frac{1+n_{\sqrt{q^2+M^2}-iT r\cdot\kappa}+n_{\sqrt{q^2+M^2}+iT r\cdot\kappa}}{2\sqrt{q^2+M^2}}\,,
\eeq
where we have conveniently separated the first two integrals. We note that the integrands are regular when $q\to M$. Moreover the first integrand does not have singularities arising from the Bose-Einstein distributions because, by assumption, $0<M<M_{r\cdot\kappa}$ and we have $M_{r\cdot\kappa}<\pi T$. We also note that all the integrals that enter the above formula are real.  For the first integral this is shown using
\beq
1+n_{ia}+n_{ib} & = & n_{ia}-n_{-ib}=\frac{e^{-ib}-e^{ia}}{(e^{ia}-1)(e^{-ib}-1)}=\frac{\sin((a+b)/2)}{2i\sin(a/2)\sin(b/2)}\nonumber\\
& = & \frac{1}{2i}\left(\frac{1}{\tan(a/2)}+\frac{1}{\tan(b/2)}\right)=\frac{1}{i}\frac{\sin((a+b)/2)}{\cos((a-b)/2)-\cos((a+b)/2)}\,.
\eeq
Contrary to the previous case, not all the $1$'s in (\ref{eq:eq}) lead to the zero temperature contribution, so we cannot use the same trick as above to compute $\Delta \hat J_\kappa(m^4,m^2_{\rm vac})$. However, since the latter is finite, we can compute it using any regulator. With a $3d$ cut-off, we have
\beq
\hat J(m^4)=\frac{1}{4\pi^2m^2}{\rm Im}\int_0^\Lambda dq\,\frac{q^2}{\sqrt{q^2-im^2}}
\eeq
and then, after some calculation,
\beq
\Delta \hat J_\kappa(m^4)=\frac{1}{32\pi^2}\ln\frac{m^4_{\rm vac}}{M^4} & - & \frac{1}{2M^2}\int_{q<M} \frac{1}{2\sqrt{M^2-q^2}}\frac{\sin(\sqrt{M^2-q^2}/T)}{\cos(r\cdot\kappa)-\cos(\sqrt{M^2-q^2}/T)}\nonumber\\
& & +\frac{1}{2M^2}\int_{q>M} \frac{1}{\sqrt{q^2-M^2}}\frac{e^{\sqrt{q^2-M^2}/T}\cos(r\cdot\kappa)-1}{e^{2\sqrt{q^2-M^2}/T}-2e^{\sqrt{q^2-M^2}/T}\cos(r\cdot\kappa)+1}\nonumber\\
& & -\frac{1}{2M^2}\int_q \frac{1}{\sqrt{q^2+M^2}}\frac{e^{\sqrt{q^2+M^2}/T}\cos(r\cdot\kappa)-1}{e^{2\sqrt{q^2+M^2}/T}-2e^{\sqrt{q^2+M^2}/T}\cos(r\cdot\kappa)+1}
\eeq
or equivalently
\beq
 \hat J_\kappa(m^4)=\frac{1}{32\pi^2}\ln\frac{m^4_{\rm vac}}{M^4} & - & \frac{1}{2M^2}\int_{q<M} \frac{1}{2\sqrt{M^2-q^2}}\frac{\sin(\sqrt{M^2-q^2}/T)}{\cos(r\cdot\kappa)-\cos(\sqrt{M^2-q^2}/T)}\nonumber\\
& & -\frac{1}{2M^2}\int_{q>M} \frac{1}{2\sqrt{q^2-M^2}}\frac{\cos(r\cdot\kappa)-e^{-\sqrt{q^2-M^2}/T}}{\cos(r\cdot\kappa)-\cosh(\sqrt{q^2-M^2}/T)}\nonumber\\
& & +\frac{1}{2M^2}\int_q \frac{1}{2\sqrt{q^2+M^2}}\frac{\cos(r\cdot\kappa)-e^{-\sqrt{q^2+M^2}/T}}{\cos(r\cdot\kappa)-\cosh(\sqrt{q^2+M^2}/T)}\,.
\eeq
Finally, we will also need $J_\kappa(m^4)$ $m^4\to 0$ (which exists for $r\cdot\kappa\notin 2\pi \mathds{Z}$). Using (\ref{eq:u}), we find
\beq\label{eq:u}
\hat J_\kappa(m^4\to 0) & = & -\int_q \frac{d}{dq^2}\frac{1+n_{q-iT r\cdot\kappa}+n_{q+iT r\cdot\kappa}}{2q}-\frac{1}{8\pi^2}+\int_q \frac{1+n_{q-iT r\cdot\kappa}+n_{q+iT r\cdot\kappa}}{4q^3}\,.
\eeq

%%%
\subsection{Sum-integral entering the potential}
The same discussion can be applied to the sum-integral 
\beq
\hat K_\kappa(m^4)\equiv\int_Q^T \ln\,(Q^4_\kappa+m^4)
\eeq 
that appears in the effective potential. At zero temperature, the integral is defined only for $m^4>0$ (again if we restrict to real values of $m^4$). We then use
\beq
\ln (Q^4+m^4)=2\,{\rm Re}\,\ln (Q^2+im^2)\,,
\eeq
together with
\beq
\int_Q \ln (Q^2+M^2)=-\frac{M^4}{32\pi^2}\left[\frac{1}{\epsilon}+\ln\frac{\bar\mu^2}{M^2}+\frac{3}{2}\right],
\eeq
valid for any non-negative $M^2$. We find
\beq\label{eq:toto6}
\hat K(m^4)\equiv\int_Q \ln (Q^4+m^4)=\frac{m^4}{16\pi^2}\left[\frac{1}{\epsilon}+\frac{1}{2}\ln\frac{\bar\mu^4}{m^4}+\frac{3}{2}\right].
\eeq
Similarly, at finite temperature, we have
\beq
\hat K_\kappa(m^4) & = & \int_q\,2{\rm Re}\Big[\sqrt{q^2+im^2}+T\ln(e^{-2\sqrt{q^2+im^2}/T}-2e^{-\sqrt{q^2+im^2}/T}\cos(r\cdot\kappa)+1)\Big],
\eeq
for $m^4>0$. In this case the first term inside the bracket corresponds to the zero temperature contribution and can be replaced by the explicit formula (\ref{eq:toto6}). Then
\beq
\Delta\hat K_\kappa(m^4;m^4_{\rm vac}) & \equiv & \hat K_\kappa(m^4)-m^4 \hat J_{T=0}(m^4_{\rm vac})\nonumber\\
& = & \frac{m^4}{32\pi^2}\left[\ln\frac{m^4_{\rm vac}}{m^4}+1\right]\nonumber\\
& + & T\int_q\,\ln(e^{-2\sqrt{q^2+im^2}/T}-2e^{-\sqrt{q^2+im^2}/T}\cos(r\cdot\kappa)+1)\nonumber\\
& + & T\int_q\,\ln(e^{-2\sqrt{q^2-im^2}/T}-2e^{-\sqrt{q^2-im^2}/T}\cos(r\cdot\kappa)+1)\,. 
\eeq
Instead, if $-M^4_{r\cdot\kappa}<m^4=-M^2<0$, we find
\beq
\hat K_\kappa(m^4) & = & \int_{q<M}\,\Big[i\sqrt{M^2-q^2}+T\ln(e^{-2i\sqrt{M^2-q^2}/T}-2e^{-i\sqrt{M^2-q^2}/T}\cos(r\cdot\kappa)+1)\Big]\nonumber\\
& + & \int_{q>M}\,\Big[\sqrt{q^2-M^2}+T\ln(e^{-2\sqrt{q^2-M^2}/T}-2e^{-\sqrt{q^2-M^2}/T}\cos(r\cdot\kappa)+1)\Big]\nonumber\\
& + & \int_q\,\Big[\sqrt{q^2+M^2}+T\ln(e^{-2\sqrt{q^2+M^2}/T}-2e^{-\sqrt{q^2+M^2}/T}\cos(r\cdot\kappa)+1)\Big].
\eeq
We have
\beq
e^{-2i\sqrt{M^2-q^2}/T}-2e^{-i\sqrt{M^2-q^2}/T}\cos(r\cdot\kappa)+1=\,2e^{-i\sqrt{M^2-q^2}/T}(\cos(\sqrt{M^2-q^2}/T)-\cos(r\cdot\kappa))\,.
\eeq
Since $0<M/T<\pi$, we can apply the formula $\ln (ab)=\ln a+\ln b$ and then
\beq
\hat K_\kappa(m^4) & = & \int_{q<M}\,T\ln(2\cos(\sqrt{M^2-q^2}/T)-2\cos(r\cdot\kappa))\nonumber\\
& + & \int_{q>M}\,\Big[\sqrt{q^2-M^2}+T\ln(e^{-2\sqrt{q^2-M^2}/T}-2e^{-\sqrt{q^2-M^2}/T}\cos(r\cdot\kappa)+1)\Big]\nonumber\\
& + & \int_q\,\Big[\sqrt{q^2+M^2}+T\ln(e^{-2\sqrt{q^2+M^2}/T}-2e^{-\sqrt{q^2+M^2}/T}\cos(r\cdot\kappa)+1)\Big],
\eeq
where each integral is real. Once again, in this case, the zero-temperature contribution is not so easily extracted and we cannot use the same trick as above to compute $\Delta_\kappa K_\kappa(m^4,m^4_{\rm vac})$. However, up to quartic divergence (that does not depend on $T$ or $r$), we can compute it using any regulator. We use a $3d$ cut-off and find
\beq
& & \Delta \hat K_\kappa(m^4,m^4_{\rm vac})\nonumber\\
& & =-\,\frac{M^4}{32\pi^2}\left(1+\ln\left(\frac{m^4_{\rm vac}}{M^4}\right)\right)+\int_{q<M}\,T\ln(2\cos(\sqrt{M^2-q^2}/T)-2\cos(r\cdot\kappa))\nonumber\\
& & +\,\int_{q>M}\,T\ln(e^{-2\sqrt{q^2-M^2}/T}-2e^{-\sqrt{q^2-M^2}/T}\cos(r\cdot\kappa)+1)\nonumber\\
& & +\,\int_q\,T\ln(e^{-2\sqrt{q^2+M^2}/T}-2e^{-\sqrt{q^2+M^2}/T}\cos(r\cdot\kappa)+1)\,. 
\eeq
We check that the derivative with respect to $M^4$ gives $-\Delta\hat J_\kappa(m^4,m^2_{\rm vac})$, as it should.\\

\end{widetext}

\end{document}